\documentclass[letterpaper,twocolumn,10pt]{article}
\usepackage{usenix, graphicx, hyperref, comment, enumitem, balance, array}
\usepackage[table]{xcolor}
\graphicspath{{figures/}}
\pdfoutput=1

\usepackage[compact]{titlesec}
\newcommand{\rz}{\rotatebox[origin=c]{-90}{Z}}

\titleformat{name=\section}{\large\bfseries}{}{0pt}{\thesection\quad}

\titleformat{name=\section,numberless}{\large\bfseries}{}{0pt}{}

\titleformat{\subsection}{\bfseries}{}{0pt}{\thesubsection\quad}

\usepackage[font=small,labelfont=bf,skip=2pt]{caption}

\newcommand{\paperfontsize}{\normalsize}

\newcommand{\Lepton}{Lepton}
\newcommand{\Dropspot}{DropSpot}
\newcommand{\Dropbox}{Dropbox}
\newcommand{\LeptonCVE}{\cite{cve2016_6234}}
\newcommand{\CameraUpload}{camera upload}
\usepackage{microtype}

\begin{document}

\date{}

\title{\Large \bf The Design, Implementation, and Deployment of a
  System to Transparently Compress Hundreds of Petabytes of Image Files
  For a File-Storage Service}

\author{\rm Daniel Reiter Horn \\
Dropbox
\and
{\rm Ken Elkabany} \\
Dropbox
\and
{\rm Chris Lesniewski-Laas} \\
Dropbox
\and
{\rm Keith Winstein} \\
Stanford University    
} 

\maketitle

\section*{Abstract}

We report the design, implementation, and deployment of
\Lepton{}, a fault-tolerant system that losslessly compresses JPEG
images to 77\% of their original size on average. \Lepton{} replaces
the lowest layer of baseline JPEG compression---a Huffman code---with
a parallelized arithmetic code, so that the exact bytes of the
original JPEG file can be recovered quickly. \Lepton{} matches the compression
efficiency of the best prior work, while decoding more than nine
times faster and in a streaming manner. \Lepton{} has been released as
open-source software and has been deployed for a year on the Dropbox
file-storage backend. As of February 2017, it had compressed
more than 203~PiB of user JPEG files, saving more than 46~PiB.

\section{Introduction}

In the last decade, centrally hosted network filesystems
have grown to serve hundreds of millions of
users. These services include
Amazon Cloud Drive,
Box,
Dropbox,
Google Drive,
Microsoft OneDrive,
and SugarSync.

Commercially, these systems typically offer users a storage
quota in exchange for a flat monthly fee, or no fee at all. Meanwhile,
the cost to operate such a system increases with the amount of data
actually stored. Therefore, operators benefit from anything that
reduces the net amount of data they store.

These filesystems have become gargantuan. After less than ten years
in operation, the \Dropbox{} system contains roughly one exabyte ($\pm$ 50\%) of data, even after applying techniques such as deduplication
and zlib compression.

We report on our experience with a different technique:
format-specific transparent file compression, based on a statistical
model tuned to perform well on a large corpus.

\begin{figure}
\includegraphics[width=\columnwidth]{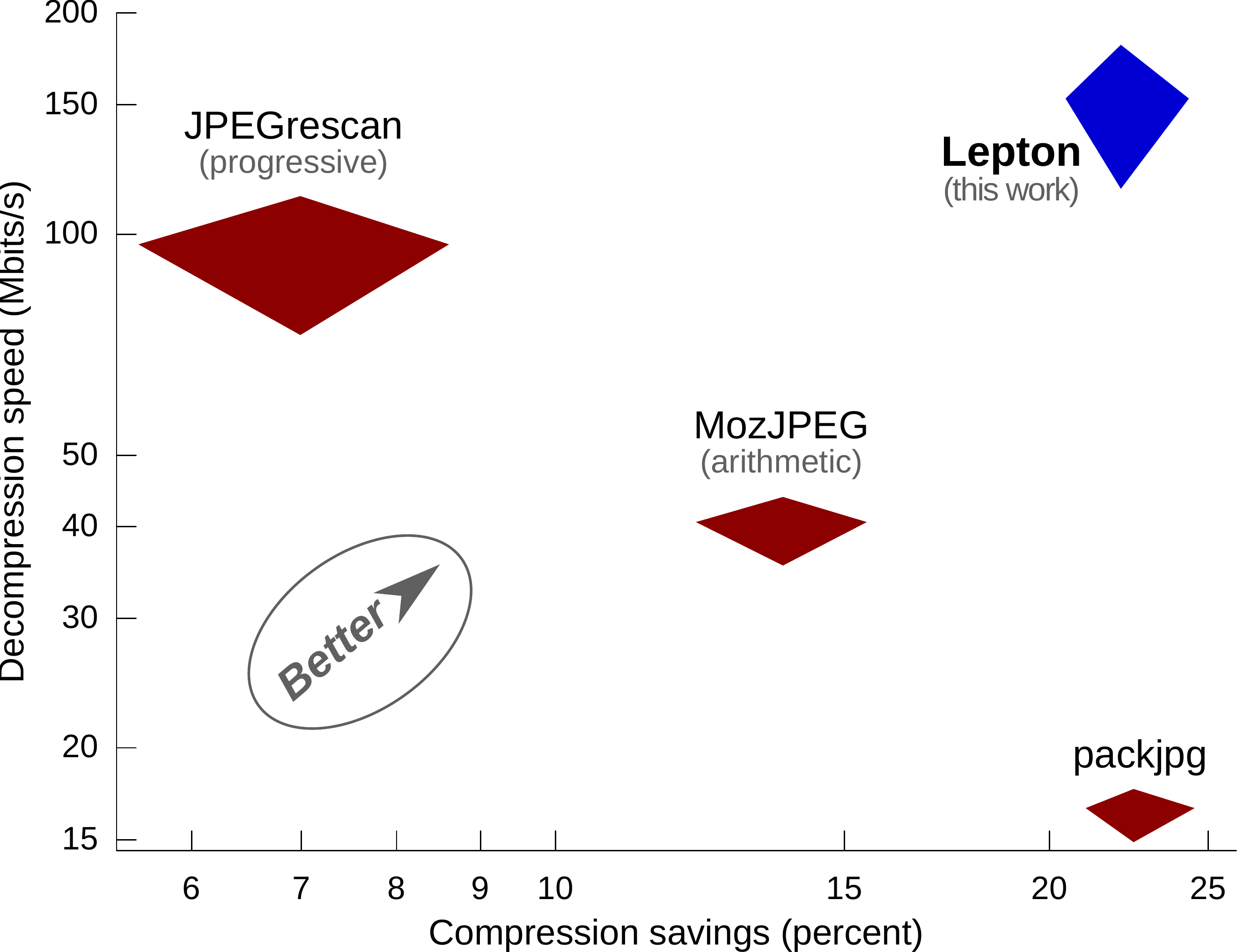}
\caption{Compression savings and decompression speed
  (time-to-last-byte)
  of four lossless JPEG compression
  tools. Diamonds show 25th, 50th, and 75th percentiles across
  200,000 JPEGs between 100 KiB and 4 MiB (\S~\ref{sec:evaluation}).}
\label{fig:marquee}
\end{figure}

In operating \Dropbox{}, we observed that JPEG
images~\cite{jpeg} make up roughly 35\% of the bytes stored. We
suspected that most of these images were compressed inefficiently,
either because they were limited to ``baseline''
methods that were royalty-free in the 1990s, or because they
were encoded by fixed-function compression chips.

In response, we built a compression tool, called
\Lepton{}, that replaces the lowest layer of baseline JPEG images---the
lossless Huffman coding---with a custom statistical model that we
tuned to perform well across a broad corpus of JPEG images stored in
\Dropbox{}. \Lepton{} losslessly compresses an average
JPEG image by about 23\%. This is expected to save \Dropbox{} more
than 8\% of its overall backend file storage. \Lepton{} introduces new techniques that allow it to match the compression savings of prior work (PackJPG) while decoding more than nine times faster and in a streaming manner (Figure~\ref{fig:marquee}).

Some of the challenges in building \Lepton{} included:

\begin{itemize}[noitemsep,topsep=0pt,parsep=0pt,partopsep=0pt]

\item \textbf{Round-trip transparency.} \Lepton{} needs to
  deterministically recover the exact bytes of the original file, even
  for intentionally malformed files, and even after updates to the
  \Lepton{} software since the file was originally compressed.

\item \textbf{Distribution across independent chunks.} The \Dropbox{}
  back-end stores files in independent 4-MiB chunks across many servers. \Lepton{} must be able to decompress any
  substring of a JPEG file, without access to other substrings.

\item \textbf{Low latency and streaming.}
  Because users are sensitive to file download latency, we must optimize
  decompression---from \Lepton{}'s format back into
  Huffman-coded JPEG output---for both time-to-first-byte and
  time-to-last-byte.
  To achieve this, the \Lepton{} format includes ``Huffman handover words''
  that enable the decoder to be multithreaded and to start transmitting bytes
  soon after a request.

\item \textbf{Security.} Before reading input data, \Lepton{} enters a restricted
  environment where the only allowed system calls are {\tt read}, {\tt write},
  {\tt exit}, and {\tt sigreturn}. \Lepton{} must pre-spawn threads and allocate
  all memory before it sees any input.

\item \textbf{Memory.} To preserve server resources, \Lepton{} must
  work row-by-row on a JPEG file, instead of decoding the entire
  file into RAM.

\end{itemize}

We deployed \Lepton{} in April 2016 on the production \Dropbox{}
service. Compression is applied immediately to all uploads of new JPEG
files.  We are gradually compressing files in existing storage and
have ramped up to a rate of roughly a petabyte per day of input,
consuming about 300 kilowatts continuously. As of February 2017,
\Lepton{} had been run on more than 150 billion user files, accounting
for more than 203~PiB of input. It reduced their size by a total of more
than 46~PiB. We have released \Lepton{} as open-source
software~\cite{leptonrepo}.

We report here on \Lepton{}'s design (\S~\ref{s:design}), evaluation
(\S~\ref{sec:evaluation}), and production deployment
(\S~\ref{s:deployment}), and share a number of case studies and
anomalies (\S~\ref{s:casestudies}) encountered in operating the system
at a large scale.

\section{Related Work}
\label{s:relwork}

In network filesystems and storage services, four general categories
of approaches are used to compress user files.

\paragraph{Generic entropy compression.} Many filesystems compress
files using generic techniques, such as the
Deflate algorithm~\cite{deflate}, which combines LZ77
compression~\cite{lz77} and Huffman coding~\cite{huffman} and 
is used by software such as zlib, pkzip, and gzip. More
recent algorithms such as LZMA~\cite{lzma},
Brotli~\cite{brotli}, and Zstandard~\cite{zstd} achieve a range of
tradeoffs of compression savings and speed.

In practice, none of these algorithms achieve much compression on
``already compressed'' files, including JPEGs. In our evaluation on
200,000 randomly selected user JPEG images, each of these algorithms
achieved savings of 1\% or less (\S~\ref{sec:evaluation}).

\paragraph{Lossy image compression.} A second category of approach
involves decoding user images and re-compressing them into a
more-efficient format, at the cost of some visual fidelity. The
``more-efficient format'' may simply be a lower-resolution or
lower-quality JPEG. The JPEG files produced by fixed-function hardware
encoders in cellular-phone cameras can typically be reduced in size by
70\% before users see a perceptual difference~\cite{tinyjpg}. The
re-compression may also use more sophisticated techniques such as WebP
or single-frame H.265. With this approach, storage savings for the
provider can be as high as users are willing to tolerate. However,
\Dropbox{} does not modify user files, precluding these approaches.

\paragraph{Format-aware pixel-exact recompression.} Several tools let users
re-compress JPEG files to achieve more-efficient compression
without affecting the decoded image. These tools include
JPEGrescan~\cite{jpgrescan} and MozJPEG~\cite{mozjpeg}, both based on
the jpegtran tool written by the Independent JPEG
Group~\cite{ijg}. The tools employ two key techniques: replacing
Huffman coding with arithmetic coding (which is more efficient but was
patent-encumbered when JPEG was formalized and is not part of the
baseline JPEG standard), and rewriting the file in ``progressive''
order, which can group similar values together and result in
more efficient coding. Xu et al.~developed a related algorithm which uses a large number of context-dependent Huffman tables to encode JPEG images, reporting average compression savings of 15\% in 30-50ms~\cite{xu2016context}. These tools preserve the exact pixels of the decoded image, but do not allow bit-exact round-trip recovery of the original file.

\paragraph{Format-aware, file-preserving recompression.} A final set of
tools can re-compress a JPEG file with round-trip recovery of the
exact bytes of the original file. These include PackJPG~\cite{packjpg,
  lakhani} and PAQ8PX~\cite{paq}. These tools use different
techniques, but in our evaluations, achieved roughly the same
compression savings (23\%) on average.

These tools use techniques unavailable in a real-time setting. For
example, one of PackJPG's compression techniques requires re-arranging
all of the compressed pixel values in the file in a globally sorted
order. This means that decompression is single-threaded, requires
access to the entire file, and requires decoding the image into RAM
before any byte can be output. The time-to-first-byte
and time-to-last-byte are too high to satisfy the service goals for the \Dropbox{}
service. PAQ8PX is considerably slower.

\Lepton{} was inspired by PackJPG and the algorithms developed by
Lakhani~\cite{lakhani}, and \Lepton{} uses the same JPEG-parsing
routines that PackJPG uses (uncmpjpg). However, unlike PackJPG,
\Lepton{} only utilizes compression techniques that can be implemented
without ``global'' operations, so that decoding can be distributed
across independent chunks and multithreaded within each chunk.

\section{\Lepton{}: Design and Implementation}
\label{s:design}

At its core, \Lepton{} is a stand-alone tool that performs round-trip
compression and decompression of baseline JPEG files. We have released
\Lepton{} as open-source software~\cite{leptonrepo}; it builds and runs
on Linux, MacOS, Windows, iOS, Android, and Emscripten
(JavaScript).

In its current deployment at \Dropbox{}, \Lepton{} is executed
directly by a back-end file server or, when a file server is under
high load, execution is ``outsourced'' from the file server to a
cluster of machines dedicated to \Lepton{} only (\S~\ref{sec:outsourcing}).
Thus, compression and decompression is currently transparent to client
software and does not reduce network utilization. In the future, we
may include \Lepton{} in the client software, in order to save network
bandwidth and distribute the computation load.

In contrast to related work (\S~\ref{s:relwork}),
\Lepton{} was designed to meet constraints specific to real-time
compression and decompression in a distributed network filesystem:

\paragraph{Distribution across independent chunks.} Decompression must
be able to be distributed across independent pieces. The \Dropbox{}
back-end stores files in chunks of at most 4~MiB, spread across many
servers. Client software retrieves each chunk
independently. Therefore, \Lepton{} must be able to decompress any
substring of a JPEG file, without access to other substrings.
By contrast, the \emph{compression} process is not subject to the same
constraints, because performance does not affect user-visible
latency. In practice, \Lepton{} compresses the first chunk in a file
immediately when uploaded, and compresses subsequent chunks later
after assembling the whole file in one place.

\paragraph{Within chunks, parallel decoding and streaming.}
Decoding occurs when a client asks to retrieve a chunk from
a network file server, typically over a consumer Internet
connection. Therefore, it is not sufficient for \Lepton{} simply to
have a reasonable time-to-last-byte for decompressing a 4-MiB
chunk. The file server must start streaming bytes quickly to start
filling up the client's network connection, even if the whole chunk
has not yet been decompressed.

In addition, average decoding speed must be fast enough to saturate a
typical user's Internet connection ($>$ 100 Mbps). In practice, this
means that decoding must be multithreaded, including both the decoding
of the \Lepton{}-format compressed file (arithmetic code decoding) and
the re-encoding of the user's Huffman-coded baseline-JPEG file. To
accomplish the latter, the \Lepton{}-format files are partitioned into
segments (one for each decoding thread), and each thread's segment
starts with a ``Huffman handover word'' to allow that thread's Huffman
encoder to resume in mid-symbol at a byte boundary.

We now give an overview of JPEG compression and discuss the
design of \Lepton{} subject to these requirements.

\subsection{Overview of JPEG compression}

A baseline JPEG image file has two sections---headers (including
comments) and the image data itself (the ``scan''). \Lepton{}
compresses the headers with existing lossless techniques
(\cite{deflate}). The ``scan'' encodes an array of quantized
coefficients, grouped into sets of 64 coefficients known as
``blocks.''
Each coefficient represents the
amplitude of a particular 8x8 basis function; to decode the JPEG
itself, these basis functions are summed, weighted by each
coefficient, and the resulting image is displayed. This is known as an
inverse Discrete Cosine Transform.

In a baseline JPEG file, the coefficients are written using a Huffman
code~\cite{huffman}, which allows more-probable values to consume
fewer bits in the output, saving space overall. The Huffman
``tables,'' given in the header, define the probability model that
determines which coefficient values will be considered more or less
probable. The more accurate the model, the smaller
the resulting file.

\Lepton{} makes two major changes to this scheme. First, it replaces
the Huffman code with an arithmetic code\footnote{\Lepton{} implements a modified version
of a VP8~\cite{rfc6386} range coder.}, a more efficient technique
that was patent-encumbered at the time the JPEG specification was
published (but no longer).
Second, \Lepton{} uses a sophisticated adaptive
probability model that we developed by testing on a large corpus of
images in the wild. The goal of the model is to produce the most accurate
predictions for each coefficient's value, and therefore the smallest file size.

\subsection{\Lepton{}'s probability model: no sorting, but more complexity}
Arithmetic probability models typically use an array of ``statistic
bins,'' each of which tracks the probability of a ``one'' vs.~a
``zero'' bit given a particular prior context. The JPEG specification
includes extensions for arithmetic-coded files~\cite{jpeg}, using a
probability model with about 300 bins.\footnote{Performance is shown
  in Figure~\ref{fig:marquee} in the diamond labeled ``MozJPEG
  (arithmetic).'' Arithmetic-coded JPEGs are not included in the
  widely-supported ``baseline'' version of the specification because
  they were patent-encumbered at the time the standard was published.} The PackJPG tool uses about 6,400 bins, after
sorting every coefficient in the image to place correlated
coefficients in the same context.

In designing \Lepton{}, we needed to avoid global operations (such
as sorting) that defeat streaming or multithreaded decoding. One key
insight is that such operations can be avoided by expanding the
statistical model to cover correlations across long distances in the
file, without needing to sort the data. \Lepton{}'s model uses 721,564
bins, each applied in a different context.

These contexts include the type of coefficient, e.g. ``DC''
(the basis function that represents the average brightness or color
over an 8x8 block) vs.~``AC,'' and the index of an ``AC'' coefficient
within a block. Each coefficient is encoded with an Exp-Golomb
code~\cite{teuhola1978compression}, and statistic bins are then used
to track the likelihood of a ``one'' bit in this encoding, given the
values of already-encoded coefficients that may be correlated.

At the start of an encoding or decoding thread, the statistic bins are
each initialized to a 50-50 probability of zeros vs.~ones. The
probabilities are then adapted as the file is decoded, with each bin counting the
number of ``ones'' and ``zeroes'' encountered so far.

The bins are independent, so a ``one'' seen in one context will not
affect the prediction made in another. As a result, the number
and arrangement of bins is important: compression efficiency suffers from the curse of dimensionality
if too many bins are used, because the coder/decoder cannot learn
useful information from similar contexts.

\subsection{Details of Lepton's probability model}
\label{sec:dct_coefficient_priors}

We developed \Lepton{}'s probability
model empirically,
based on a handful of photos that we captured with popular
consumer cameras. We then froze the model and tested on randomly
selected images from the \Dropbox{} filesystem; performance on the
``development'' images correlated well with real-world performance
(\S~\ref{sec:evaluation}). The development images and the full
probability model are included in the open-source
release~\cite{leptonrepo} and are detailed in
Appendix~\ref{sec:bin_index}. We briefly summarize here.

For each 8x8 JPEG block, \Lepton{} encodes 49 AC coefficients (7x7), 14
``edge'' AC coefficients of horizontal (7x1) and vertical (1x7) variation, and
1 DC coefficient.

For the 7x7 AC coefficients, we predict the Golomb code bits by averaging the
corresponding coefficients in the \emph{above}, \emph{left}, and
\emph{above-left} blocks.
Hence, the bins for bits of $C_i$ are indexed by
$\langle i, \lfloor \log_2 ( |A_i| + |L_i| + \frac{1}{2} |AL_i| ) \rfloor \rangle $.

For the 7x1 and 1x7 AC coefficients, we use the intuition supplied by
Lakhani~\cite{lakhani} to transform an entire column of a two-dimensional DCT
into a one-dimensional DCT of an edge row.
In this manner we can get pixel-adjacent 1D DCT coefficients from the
bottom-most row of the \emph{above} block and the top row of the current block.
Likewise, we can use the neighboring right-most column of the \emph{left} block
to predict the left-most 1D DCT column of the current block.

To predict the DC coefficient, we assume image gradients are smooth
across blocks.
Linearly extrapolating the last two rows of pixels of the \emph{above} and \emph{left} blocks
yields 16 edge pixel values.
Since the DC coefficient is decoded last, we can use every AC coefficient to
compute a predicted DC offset which minimizes average differences between the
decoded block's edge pixels and the edges extrapolated from neighbors.
We only encode the delta between our predicted DC value and the true DC value,
so close predictions yield small outputs.
We achieved additional gains by
indexing the statistics bins by outlier values and the variance of edge pixels, enabling \Lepton{}'s model
to adapt to non-smooth gradients.

These techniques yield significant improvements over using the
same encoding for all coefficients (\S~\ref{sec:technique_breakdown}).

\subsection{Decompressing independent chunks, with multithreaded output}

When a client requests a chunk from the \Dropbox{} filesystem, the
back-end file servers must run \Lepton{} to decode the compressed
chunk back into the original JPEG-format bytes. Conceptually this
requires two steps: \emph{decoding} the arithmetic-coded coefficients
(using the \Lepton{} probability model) and then \emph{encoding} the
coefficients using a Huffman code, using the Huffman probability model
given in the file headers.

The first step, arithmetic decoding, can be parallelized by
splitting the coefficients into independent segments, with each segment
decoded by one thread. Because the \Lepton{} file format is under our
control, we can use any number of such
segments. However, adding threads decreases compression
  savings, because each thread's
  model starts with 50-50 probabilities and adapts independently.

The second step, Huffman encoding, is more challenging. The
user's original JPEG file is not under our control and was not
designed to make multithreaded encoding possible. This step
can consume a considerable amount of CPU resources in the critical
path and would consume a large fraction of the total latency if not parallelized.
Moreover, \Dropbox{}'s 4-MiB chunks may split the JPEG
file arbitrarily, including in the middle of a Huffman-coded
symbol. This presents a challenge for distributed decoding of
independent chunks.

To solve this, we modified the \Lepton{} file format to include
explicit ``Huffman handover words'' at chunk and thread
boundaries. This represents state necessary for the JPEG
writer to resume in the middle of a file, including in
mid-symbol. In particular, the Huffman handover words include the
previous DC coefficient value (16 bits), because DC values are encoded
in the JPEG specification as deltas to the previous DC value,
making each chunk dependent on the previous. They also include the bit
alignment or offset and partial byte to be written.

The Huffman handover words allow decoding to be parallelized both
across segments within a chunk, and across chunks distributed across
different file servers. Within a chunk, the Huffman handover words
allow separate threads to each write their own segment of the JPEG
output, which can simply be concatenated and sent to the user. The
\Lepton{} file format also includes a Huffman handover word and the
original Huffman probability model at the start of each chunk, allowing
chunks to be retrieved and decoded independently.

\section{Performance evaluation}
\label{sec:evaluation}

For our benchmarks, we collected 233,376 randomly sampled data chunks beginning
with the JPEG start-of-image marker (0xFF, 0xD8) from the \Dropbox{} store.
Some of these chunks
are JPEG files, some are not JPEGs, and some are the first 4 MiB of a large
JPEG file.
Since \Lepton{} in production is applied on a chunk-by-chunk basis,
and 85\% of image storage is occupied by chunks with the JPEG start-of-image marker,
this sampling gives a good approximation to the deployed system.

\Lepton{} successfully compresses 96.4\% of the sampled chunks.
The remaining 3.6\% of chunks (accounting for only 1.2\% of bytes) were
non-JPEG files, or JPEGs not supported by \Lepton{}.
\Lepton{} detects and skips these files.

\subsection{Size and speed versus other algorithms}

\begin{figure}
  \includegraphics[width=7.95cm]{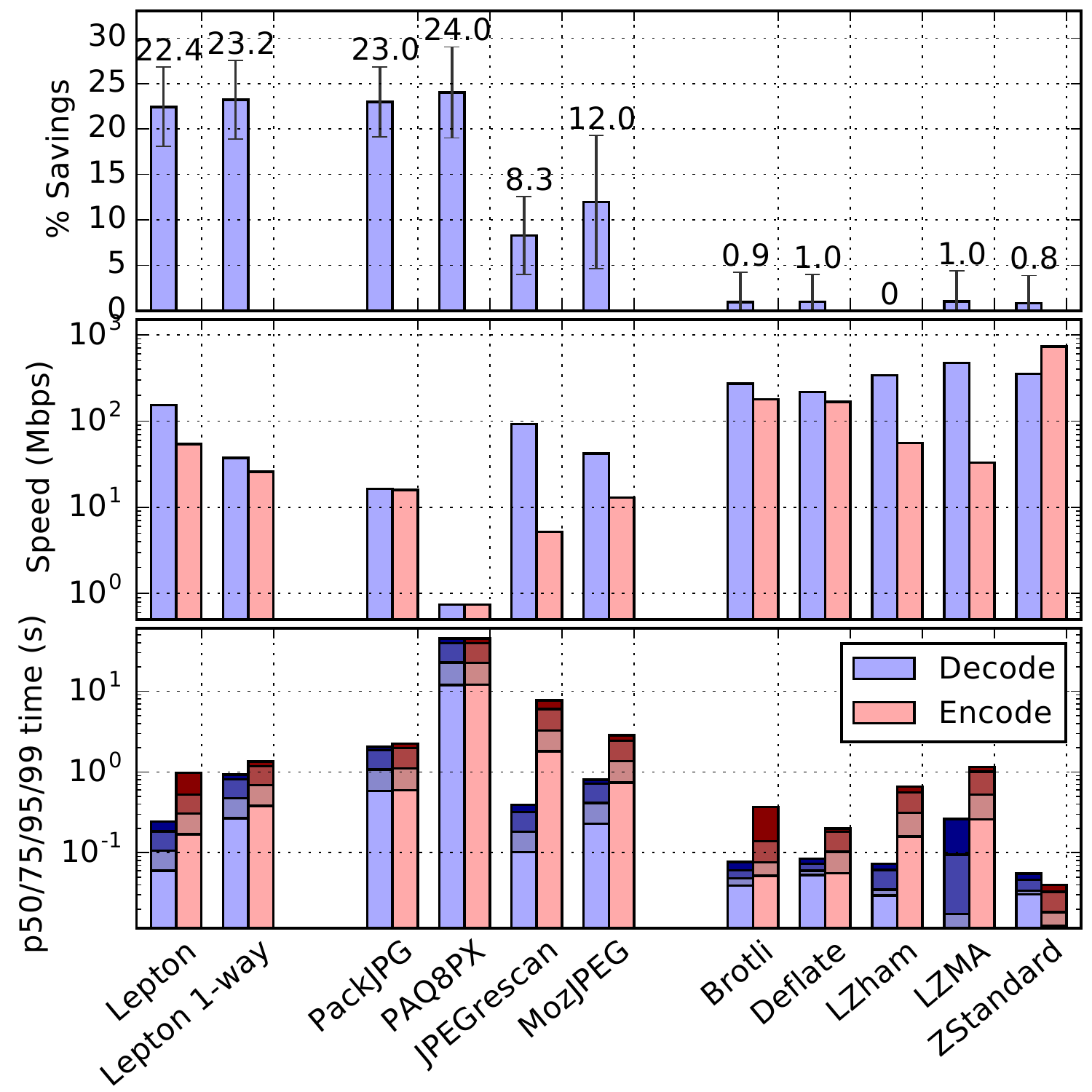}
  \caption{Compression savings and speed of codecs on the
  benchmark data-set (including chunks that \Lepton{} cannot compress).
  Generic codecs (right) are fast, but only able to compress the JPEG
  header. JPEG-aware codecs (center) compress well, but are slow.
  \Lepton{} (far left) is fast and compresses well.}
  \label{fig:compression_comparison}
\end{figure}

In Figure~\ref{fig:compression_comparison} we compare \Lepton{}'s performance
against other compression algorithms built with the Intel C++ Compiler 16.0 (icc) on a 2.6 GHz Intel Xeon E5 2650v2.
We used the full benchmark data-set, including chunks that \Lepton{} rejects as corrupt, progressive, or in the CMYK color space.

\Lepton{} is the fastest of any format-aware compression
algorithm, and it compresses about as well as the best-in-class algorithms.
We also evaluated a single-threaded version of \Lepton{} (\Lepton{} 1-way), which we modified for maximum compression savings, by tallying statistic bins across the whole image rather than independent thread-segments.
The format-aware PAQ8PX algorithm edges out single-threaded \Lepton{}'s
compression ratio by 0.8 percentage points, because it incorporates a variety of alternative compression engines that
work on the 3.6\% of files that \Lepton{} rejects as corrupt.
However, PAQ8PX pays a price in speed: it encodes 35 times slower and decodes
50 times slower than single-threaded \Lepton{}.%

In production, we care about utilizing our CPU, network and storage resources
efficiently (\S~\ref{sec:economics}), and we care about response time to users.
\Lepton{} can use 16 CPU cores to decode at 300~Mbps by
processing 2 images concurrently.
For interactive performance, we tuned \Lepton{} to decode image chunks in under
250 ms at the 99th percentile (p99), and the median (p50) decode time is under
60 ms.
This is an order of magnitude faster than PackJPG, $1.5\times$--$4\times$ faster than
JPEGrescan and MozJPEG, and close to Deflate or Brotli.
Encoding is also fast: 1~s at the 99th percentile and 170~ms in the median case, substantially better
than any other algorithm that achieves appreciable compression.

\subsection{Memory usage}

\Lepton{} shares production hosts with other, memory-hungry processes,
so limiting memory usage was a significant design goal.
This is particularly important for the decode path because, under memory
pressure, our servers can fall back to encoding using Deflate, but we do not
have that option for decoding.

\begin{figure}
  \includegraphics[width=7.95cm]{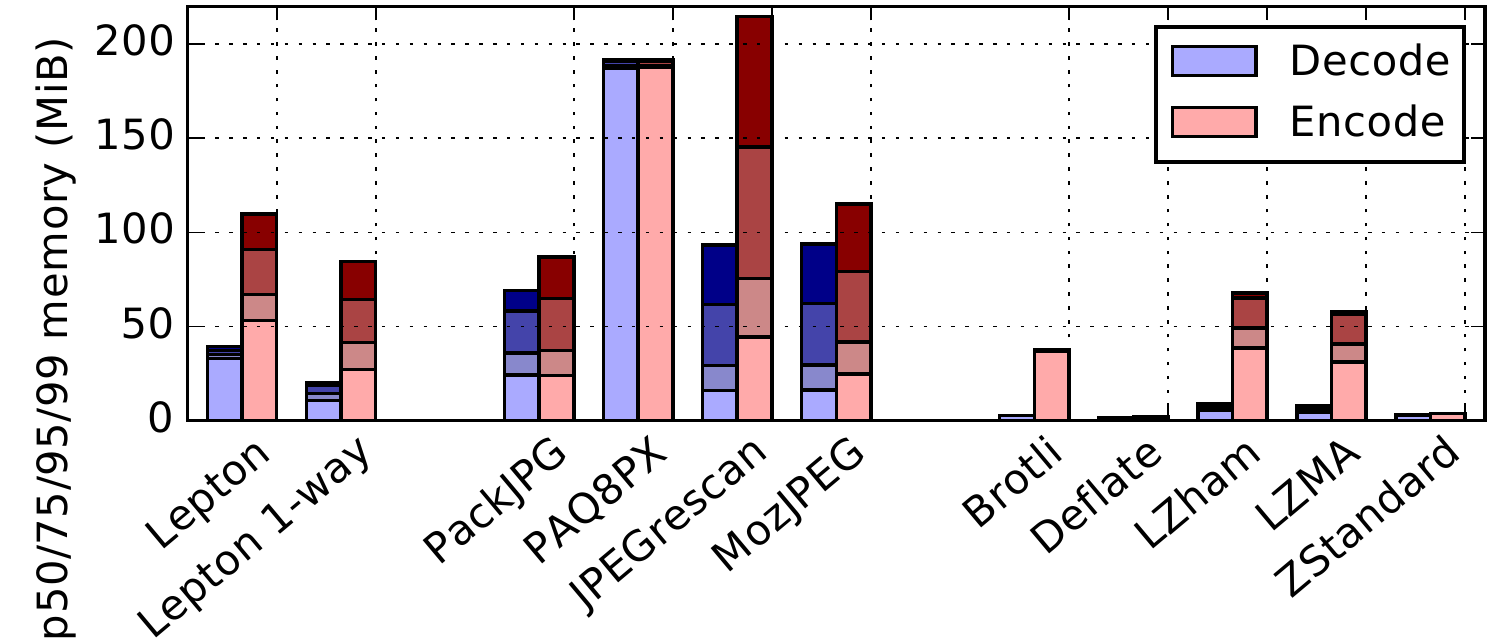}
  \caption{Max resident memory used by different algorithms.}
  \label{fig:compression_mem}
\end{figure}

Figure~\ref{fig:compression_mem} shows the memory usage of \Lepton{} and other
algorithms on our benchmark.
For decoding, single-threaded \Lepton{} uses a hard maximum of 24 MiB
to store the model and temporary buffers.
Multithreaded \Lepton{} duplicates the model for each thread, using 39 MiB at the
99th percentile.
This compares favorably with other algorithms using 69--192 MiB.

Decoding can stream the output, but encoding currently retains all pixels in
memory so each thread can operate on its own spatially contiguous region of pixels.
Thus, the memory profile of \Lepton{} encoding is similar to PackJPG and
MozJPEG.

\subsection{Compression savings by component}
\label{sec:technique_breakdown}
\begin{figure}
  \small
  \hfuzz=20pt  
  \begin{tabular}{@{}llll@{}}
    {\footnotesize Category }& {\footnotesize Original bytes } & {\footnotesize Compression Ratio \paperfontsize{}} & {\footnotesize Bytes saved \paperfontsize{}} \\
    \hline
    Header         &  2.3\% \small $\pm$ 4.2 & 47.6\% \small $\pm$ 19.8 &  1.0\% \small $\pm$ 1.8 \\
    7x7 AC         & 49.7\% \small $\pm$ 7.1 & 80.2\% \small $\pm$  3.2 &  9.8\% \small $\pm$ 1.7 \\
    7x1/1x7        & 39.8\% \small $\pm$ 4.7 & 78.7\% \small $\pm$  3.9 &  8.6\% \small $\pm$ 2.2 \\
    DC             &  8.2\% \small $\pm$ 2.6 & 59.9\% \small $\pm$  8.7 &  3.4\% \small $\pm$ 1.6 \\
    \hline
    Total          & 100\%               & 77.3\% \small $\pm$  3.6 & 22.7\% \small $\pm$ 3.6 \\
  \end{tabular}
  \paperfontsize{}
  \caption{Breakdown of compression ratio (compressed size / uncompressed size) by JPEG file components.}
  \label{fig:breakdown}
\end{figure}

For the sampled chunks that \Lepton{} successfully compresses,
Figure~\ref{fig:breakdown} shows how each part of the file contributed to the
total compression ratio, among JPEG files, of 77.3\% $\pm$ 3.6 (with multithreading enabled).

Lakhani-inspired edge prediction~\cite{lakhani} contributes 1.5\% of overall
compression savings.
Compared with baseline PackJPG~\cite{packjpg} (which used the same predictions
for all AC coefficients), it improved the compression of 7x1/1x7 AC
coefficients from 82.5\% to 78.7\%.
DC gradient prediction 
contributes 1.6\% of overall savings, improving the compression of DC
coefficients from 79.4\% (using baseline PackJPG's approach) to 59.9\%.
\footnote{%
By ``baseline PackJPG'', we refer here to the algorithm described in the 2007
publication~\cite{packjpg}.
However, for fairness, all other performance comparisons in this paper
(e.g.,
Figures~\ref{fig:marquee},~\ref{fig:compression_comparison},~\ref{fig:compression_mem})
use the latest version of the PackJPG software, which has various unpublished
improvements over the 2007 version.%
}

%

\section{Deployment at Scale}
\label{s:deployment}

To deploy \Lepton{} at large scale, without compromising durability, we faced two key design requirements: determinism and security. Our threat model includes intentionally corrupt files that seek to compress or decompress improperly
or cause \Lepton{} to execute unintended or arbitrary code or otherwise consume excess resources.

With determinism, a single successful roundtrip test guarantees that
the file will be recoverable later. However, it is difficult to prove
highly optimized C++ code to be either deterministic or secure, even
with bounds checks enabled. Undefined behavior is a core mechanism by
which C++ compilers produce efficient code~\cite{undefined-behavior},
and inline assembly may be required to produce fast inner loops, but
both hinder analysis of safety and determinism. At present, safer languages
(including Rust and Java) have difficulty achieving high performance
in image processing without resorting to similarly unsafe mechanisms.

We wrote \Lepton{} in C++, and
we enforce security and determinism using Linux's secure computing mode ({\tt
SECCOMP}). We have also cross-tested \Lepton{} at scale using multiple
different compilers.

\subsection{Security with SECCOMP}
When {\tt SECCOMP} is activated, the kernel disallows all system calls a process may make except for \texttt{read}, \texttt{write}, \texttt{exit} and \texttt{sigreturn}.  This means a program may not open new files, fork, or allocate memory.  \Lepton{} allocates a zeroed 200-MiB region of memory upfront, before reading user data, and sets up pipes to each of its threads before initiating {\tt SECCOMP}. Memory is allocated from the main thread to avoid the need for thread synchronization.

\subsection{Imperfect efforts to achieve deterministic C++}
To help determinism, the \Lepton{} binary is statically linked, and all heap
allocations are zeroed before use.  However, this setup was insufficient to
detect a buffer overrun from incorrect index math in the \Lepton{} model
(\S~\ref{sec:reversedindices}).  We had an additional fail-safe mechanism to
detect nondeterminism.  Before deploying any version of \Lepton{}, we run it on
over a billion randomly selected images (4 billion for the first version), and
then decompress each with the same binary and additionally with a single-threaded
version of the same code built with gcc using the address sanitizer
and undefined-behavior checker~\cite{asan}. This system detected the
nondeterministic buffer overrun after just a few million images were processed
and has caught some further issues since
(\S~\ref{sec:false_alarms}).

\subsection{Impact}
As of Feb.~16, 2017, \Lepton{} has encoded 203 PiB of images, reducing the size of those images by 46 PiB. The traffic has been divided between live encode traffic and a steady rate of background encoding of older images. \Dropbox{} has decoded and served 427 billion \Lepton{}-compressed images.

\subsection{Workload}
\begin{figure}
\includegraphics[width=7.95cm]{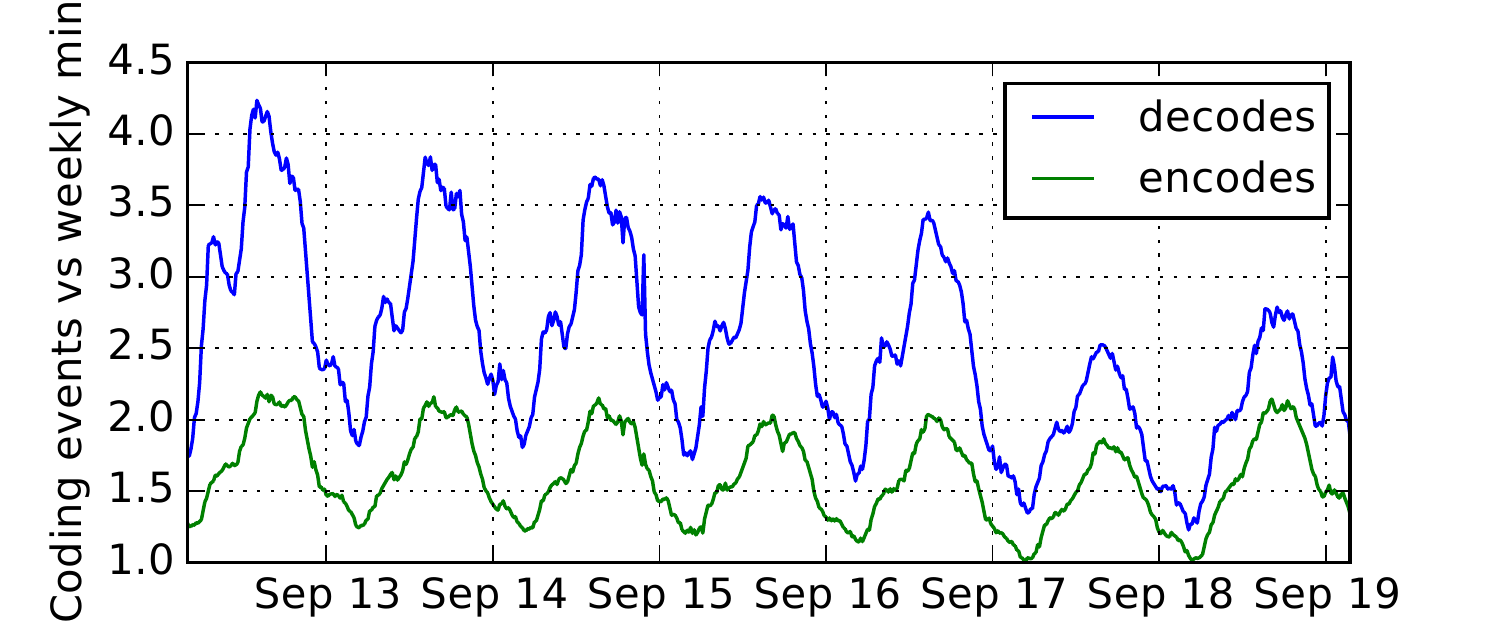}
\caption{Weekday upload rates are similar to weekends, but weekday download
  rates of \Lepton{} images are higher.}
\label{fig:absolute_decodes}
\end{figure}

\Lepton{} has been in production since April 14\footnote{All dates are in 2016 and times are
  in UTC.}, and has been on the serving path for all uploads and hundreds of billions of downloads.  A typical week can be observed in Figure \ref{fig:absolute_decodes}. On the weekends, users tend to produce the same number of photos but sync fewer to their clients, so the ratio of decodes to encodes approaches 1.0.  On weekdays, users tend to consume significantly more photos than they produce and the ratio approaches 1.5. Over the last 6 months \Dropbox{} has encoded images with \Lepton{} at between 2 and 12~GiB per second.

\begin{figure}
\includegraphics[width=7.95cm]{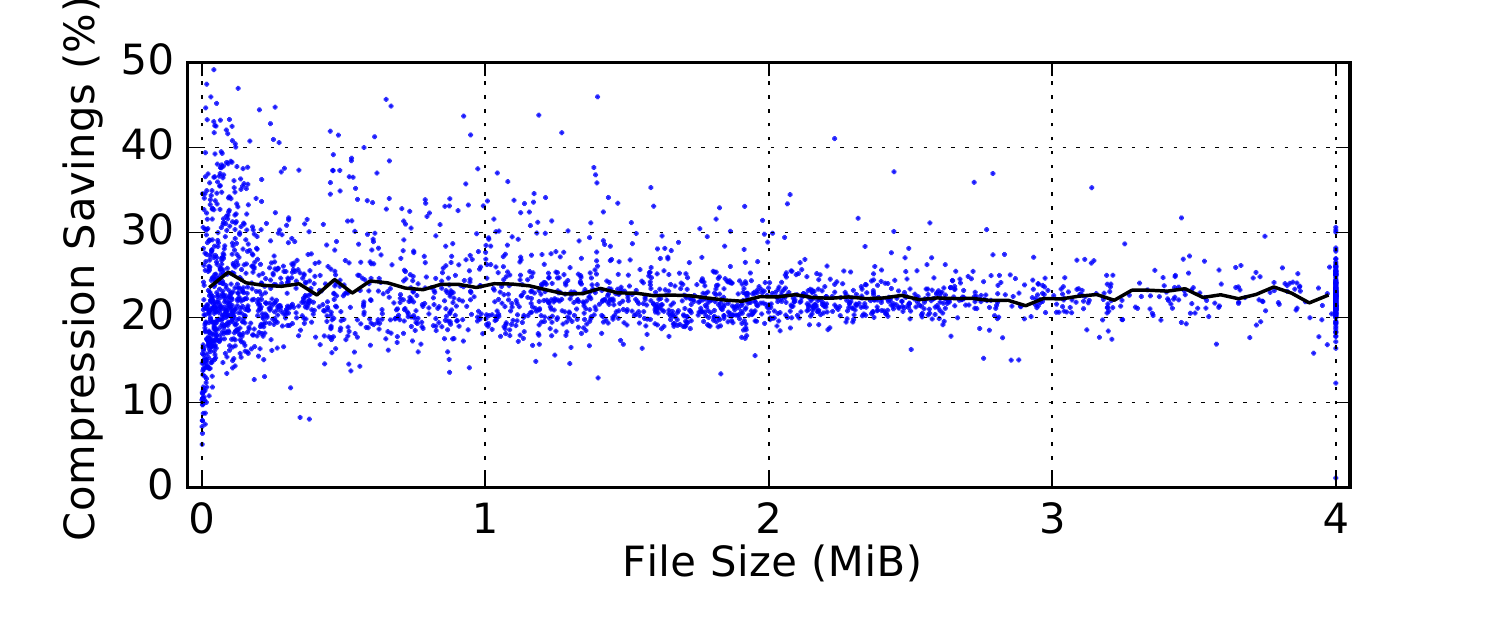}
\caption{Compression savings on production workloads are uniform across file sizes.}
\label{fig:real_world_ratio}
\end{figure}

When the \Lepton{} encoder accepts a three-color, valid, baseline JPEG, that file compresses down to, on average, 77.31\% of its original size (22.69\% savings). The savings are uniform across file sizes, as illustrated in Figure~\ref{fig:real_world_ratio}.

\begin{figure}
\includegraphics[width=7.95cm]{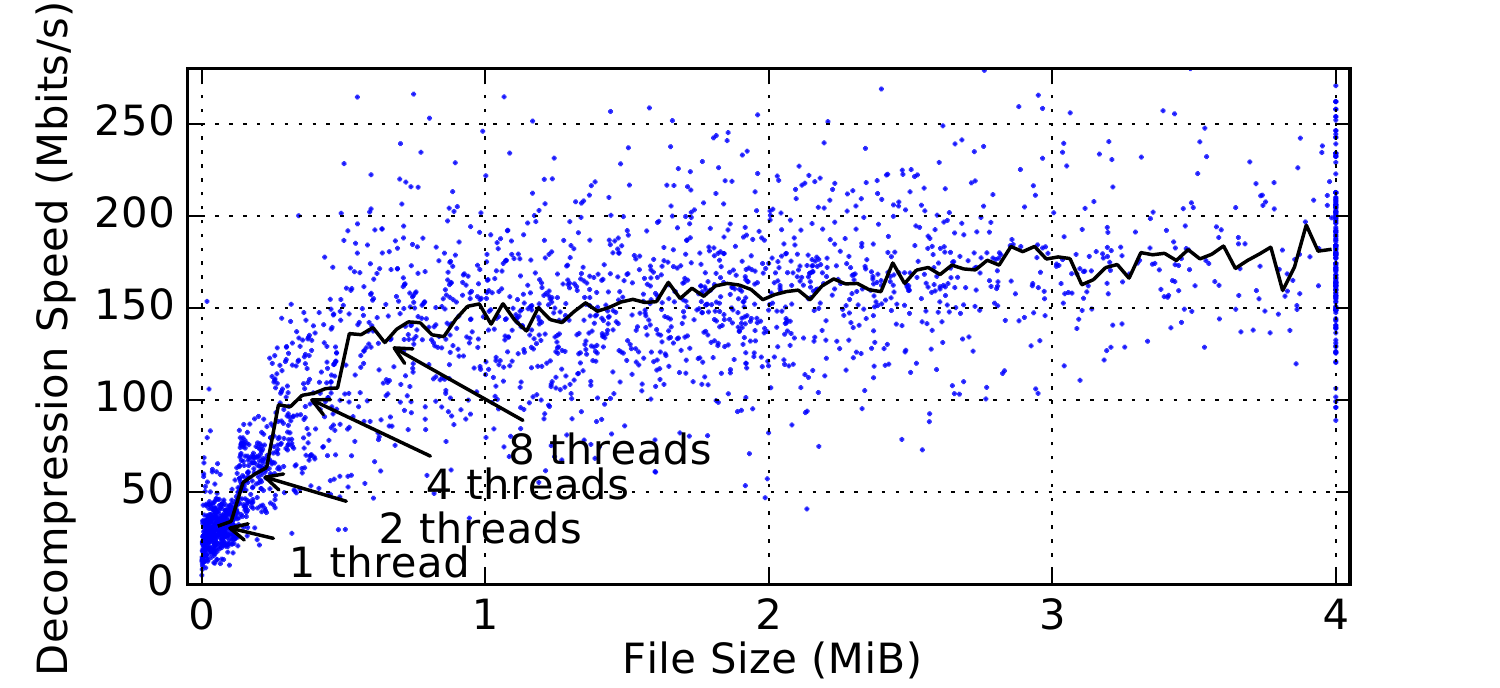}
\caption{Decompression speed given an input file size.}
\label{fig:real_world_decompression_rate}
\end{figure}

Small images are able to compress well because they are configured with fewer
threads than larger images and hence have a higher proportion of the image upon
which to train each probability bin. The number of threads per image was
selected empirically based on when the overhead of thread startup outweighed
the gains of multithreading. The multithreading cutoffs can be noticed in the
production performance scatter plot in Figure
\ref{fig:real_world_decompression_rate}. Because \Lepton{} is streaming,
the working set is roughly fixed in size. Profiling the decoder using hardware
counters confirmed that the L1 cache is rarely missed. Nearly all L1 delays
are due to pipeline stalls, not cache misses.

\begin{figure}
\includegraphics[width=7.95cm]{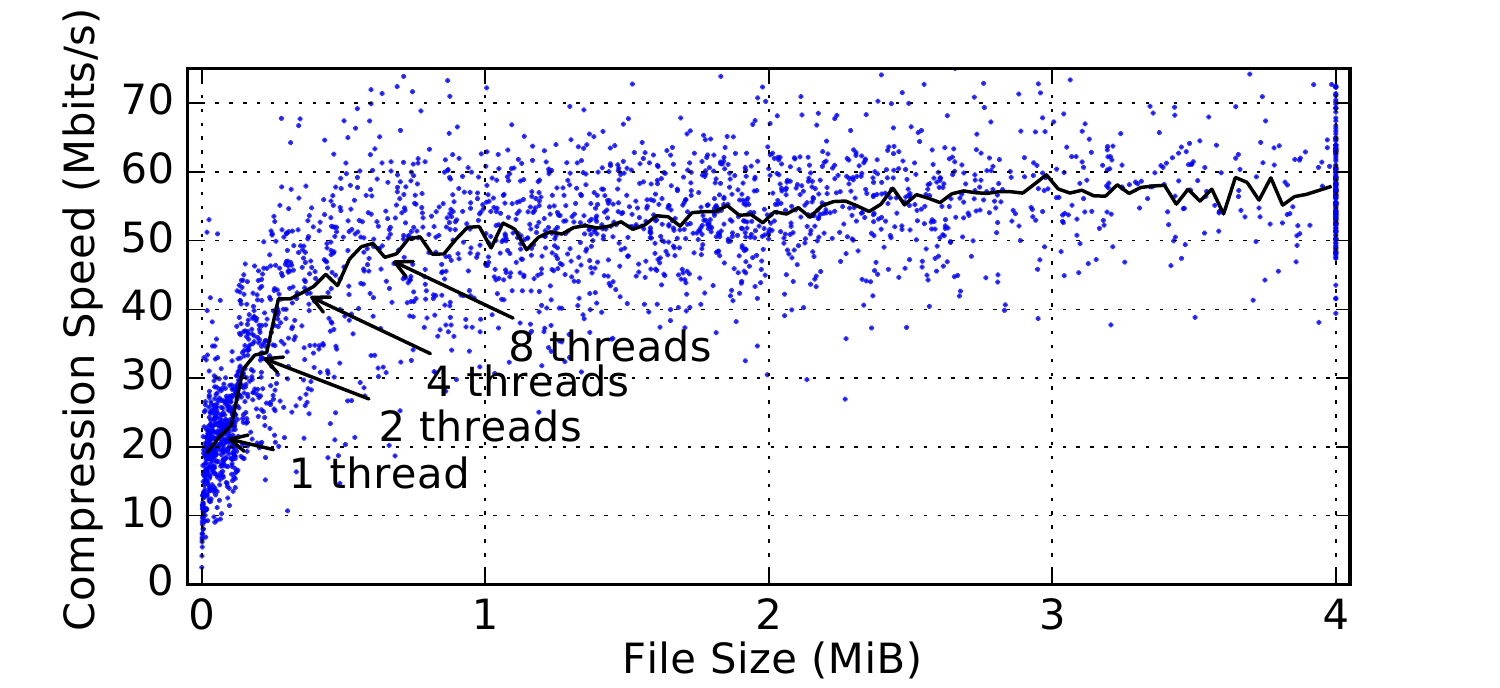}
\caption{Compression speed given an input file size.}
\label{fig:real_world_compression_rate}
\end{figure}

The compression speed, shown in Figure~\ref{fig:real_world_compression_rate},
is similar, but it is almost unaffected by the benefit of moving to 8 threads
from 4. This is because at 4 threads the bottleneck shifts to the JPEG Huffman
decoder, away from the arithmetic coding. This is solved in the \Lepton{} decoder with
the Huffman handover words, but the \Lepton{} encoder must decode the original JPEG
serially.

\subsection{Outsourcing}
\label{sec:outsourcing}

Blockservers are machines that, among other tasks, respond to requests to store
or retrieve data chunks, whether \Lepton{}-compressed JPEGs or
Deflate-compressed files. Load balancers, which do not inspect the type of
request, randomly distribute requests across the blockserver fleet.

\begin{figure}
\includegraphics[width=7.95cm]{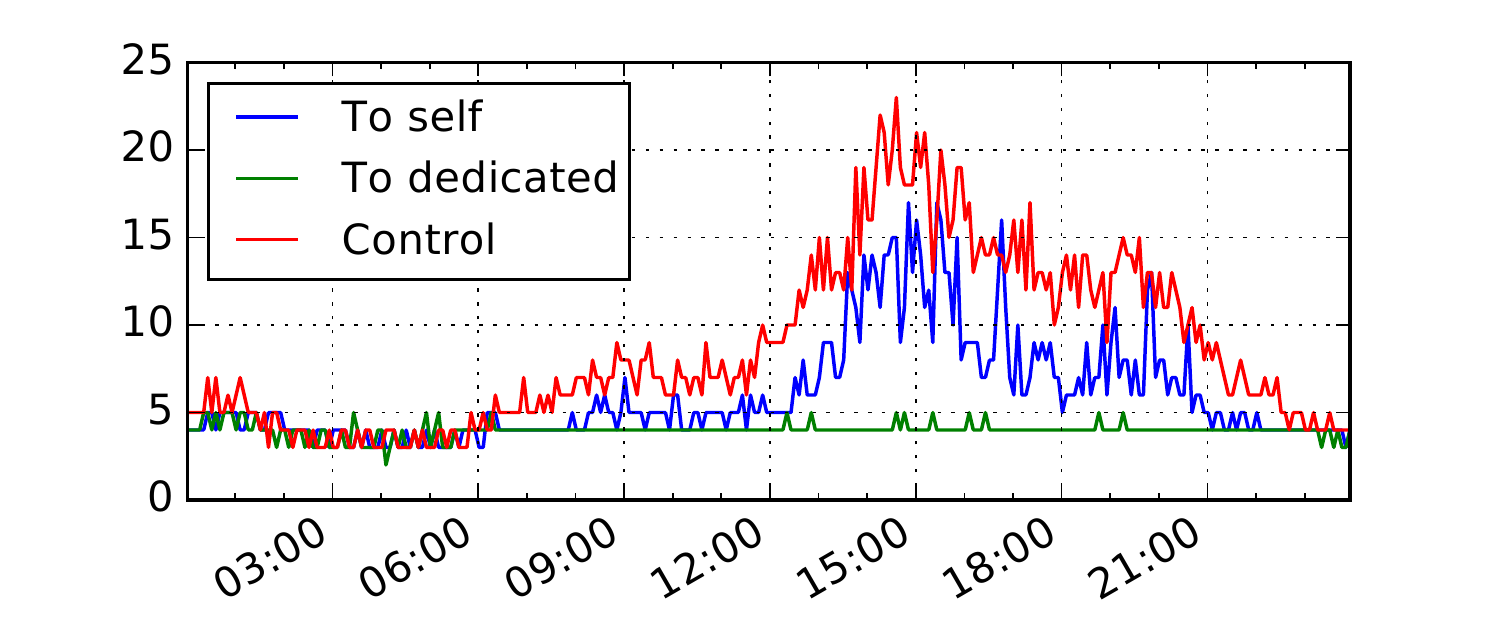}
\caption{99th-percentile of concurrent \Lepton{} processes on Sept.~15 with listed outsourcing strategy and threshold=4.}%
\label{fig:concurrency}
\end{figure}

Each blockserver has 16 cores, which means that 2 simultaneous \Lepton{}
decodes (or encodes) can completely utilize a machine. However,
blockservers are configured to handle many more than 2 simultaneous requests,
because non-\Lepton{} requests are far less resource-intensive. Therefore,
a blockserver can become oversubscribed with work, negatively affecting
\Lepton{}'s performance, if it is randomly assigned
3 or more \Lepton{} conversions at once. Without outsourcing, there are an average of 5 encodes/s during the
Thursday peak. Individual blockservers will routinely get 15 encodes
at once during peak, to the point where there is never a full minute where
there isn't at least one machine doing 11 parallel encodes during an hour of
peak traffic, as illustrated in the Control line in
Figure~\ref{fig:concurrency}.

We mitigated this problem by allowing overloaded blockservers to ``outsource''
compression operations to other machines. Inspired by the power of two random choices~\cite{tworandomchoices}, \Lepton{} will outsource any compression operations that occur on machines that have more than three conversions happening at a time.

Under normal operation, when the system is under low load, \Lepton{} operates by listening on a Unix-domain socket for files. A file is read from the socket, and the (de)compressed output is written back to the socket. The file is complete once the socket is shut down for writing. When outsourcing, instead of a Unix-domain-socket connection to a local \Lepton{} process, the blockserver instead will make a TCP connection to a machine tagged for outsourcing within the same building in the same datacenter.\footnote{Initially it seemed to make sense logistically to select an outsourcing destination simply in the same metro location as the busy blockserver. However, in measuring the pairwise conversion times, %
our datacenters in an East Coast U.S.~location had a 50\% latency
increase for conversions happening in a different building or room
within, and in a West Coast location, the difference could be as high as
a factor of 2.} The overhead from switching from a Unix-domain socket to
a remote TCP socket was 7.9\% on average.

We have two alternative strategies for selecting machines for outsourcing. The
simpler idea was to dedicate a cluster of machines ready to serve
\Lepton{} traffic for overloaded blockservers. This cluster is easy to
provision to meet traffic demands and can be packed full of work since
there are no contending processes on the machines.

Our other strategy was to mark each blockserver as an
outsourcing target for other blockservers (denoted ``To Self''). The intuition
is that in the unlikely case of a current blockserver being overloaded, the
randomly chosen outsource destination is likely to be less overloaded than the
current machine at the exact contended instant.

\subsubsection{Outsourcing Results}

\begin{figure}
\includegraphics[width=7.95cm]{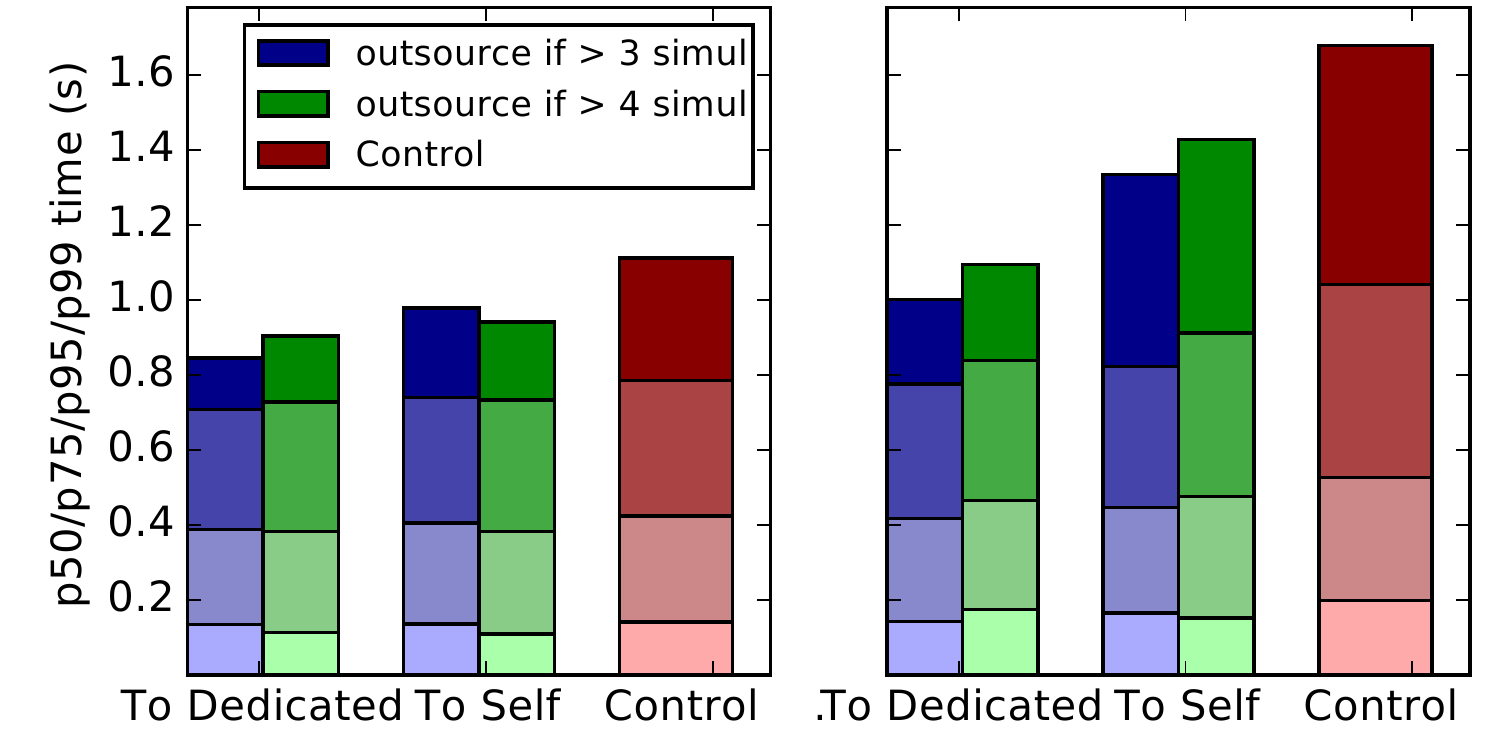}
\caption{Percentile timings of JPEG compression near to peak traffic (left) and at
  peak traffic (right) with 2 outsourcing strategies when concurrent
  \Lepton{} processes exceed a threshold (denoted with bar color) on the local machine.}%
\label{fig:outsourcing}
\end{figure}

Figure \ref{fig:outsourcing} illustrates that outsourcing reduces the
p99 by 50\% at peak from 1.63 s to 1.08 s and the p95 by 25\%.

The dedicated cluster reduces the p99 more than simply outsourcing directly to other, busy, blockservers, especially at peak. However, rebalancing traffic within the same cluster of blockservers has the added effect of reducing the p50 as well, since there are fewer hotspots because of the additional load balancing.

\subsection{Backfill}
\Lepton{} has been configured to use spare compute capacity to gradually
compress older JPEG files in storage, a process we call ``backfilling.'' To this end, we developed a small system called \Dropspot{}. \Dropspot{} monitors the spare capacity in each server room, and when the free machines in a room exceed a threshold, a machine is allocated for \Lepton{} encoding. When too few machines are free, \Dropspot{} releases some.

Wiping and reimaging the machine with the necessary software takes 2-4 hours, so a sufficiently diverse reserve of machines must be available for on-demand use.
The backfill system can be run on Amazon spot instances, but our goal of 6,000 encodes per second has been attainable using spare capacity.

In July, all user accounts were added to a sharded table in a database service backed by MySQL. For a \Lepton{} backfill worker to find images to encode, it sends a request to the metadata servers (metaservers) to request work from a randomly chosen shard. The metaserver selects the next 128 user-ids from the table in the corresponding shard. The metaserver scans the filesystem for each user, for all files with names containing the case-insensitive string ``.jp'' (likely jpeg or jpg). The metaserver builds a list of SHA-256 sums of each 4-MiB chunk of each matching file until it obtains up to 16,384 chunks. The metaserver returns a response with all the SHA-256 sums, the list of user ids to process, and a final, partial user with a token to resume that user. The worker then downloads each chunk and compresses it. It double-checks the result with the gcc address-sanitizer version of \Lepton{} in both single and multithreaded mode, and uploads the compressed version back to \Dropbox{}.

\subsubsection{Cost Effectiveness}
\label{sec:economics}
\begin{figure}
\includegraphics[width=7.95cm]{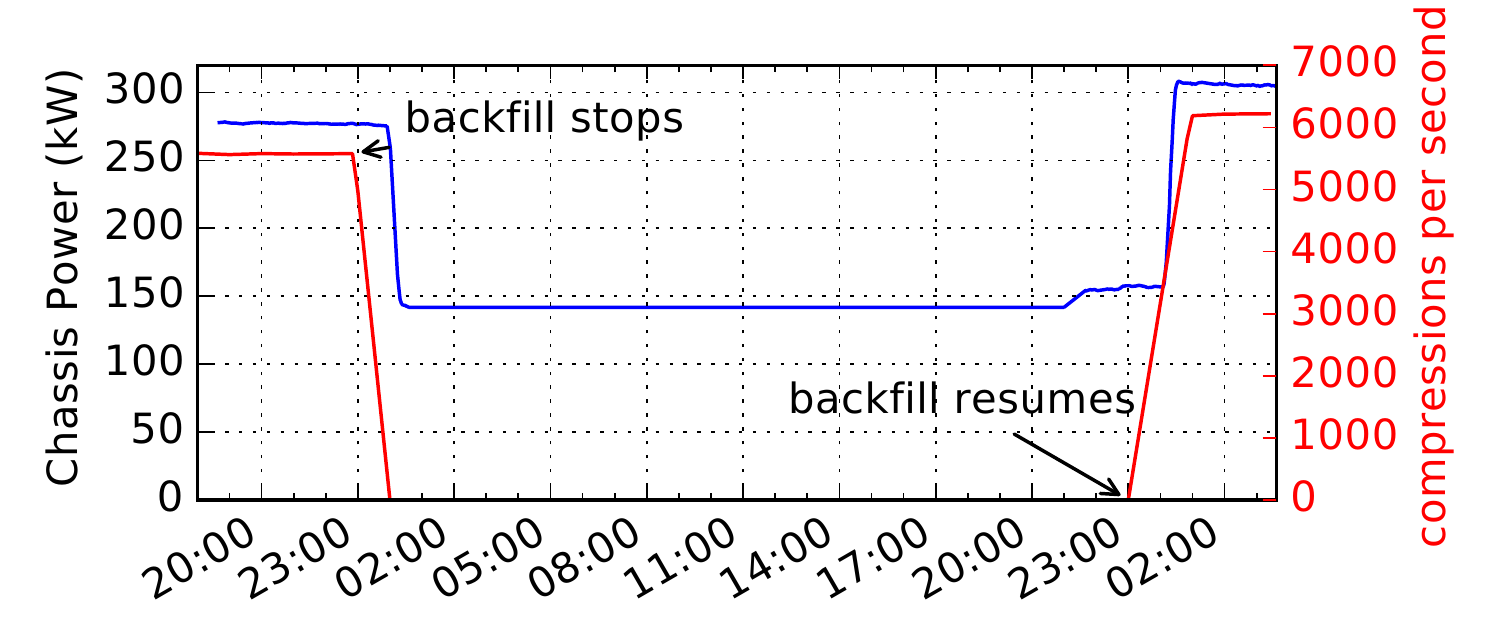}
\caption{Overall \Dropbox{} datacenter power usage graph on Sept.~26, 2016. During an outage, \Lepton{} backfill was disabled. When shut off, the power usage dropped by 121 kW.}
\label{fig:powerdrop}
\end{figure}

The cluster has a power footprint of 278~kW and it encodes 5,583 chunks per second (Figure \ref{fig:powerdrop}).
This means that one kWh can be traded for an average of 72,300 \Lepton{}
conversions of images sized at an average of 1.5 MB each.
Thus, a kWh can save 24 GiB of storage, permanently.
The power usage includes three extraneous decodes, which could be tallied as future reads, since the file is redundantly checked three times during backfill. %

Imagining a depowered 5TB hard drive costing \$120 with no redundancy or checks, the savings from the extra space would be worthwhile as long as a kWh costed less than \$0.58. Even in Denmark, where electricity costs about \$0.40 per kWh, \Lepton{} would be a worthwhile investment for somewhat balanced read/write load of 3:1. Most countries offer prices between \$0.07 and \$0.12 per kWh. With cross-zone replication, erasure coding, and regular self-checks, 24~GiB of storage costs significantly more in practice. For instance, buying a year storage for 24 GiB on Amazon S3's Infrequent Access Storage tier, as of February 2017, would cost \$3.60 each year, excluding any data read fees, making the actual savings even more clear.

To get the full 5,583 conversions per second, 964 machines are required. This means that each Intel Xeon E5 2650 v2 at 2.6~GHz can backfill 5.75~images per second. This means each server can process 181,500,000 images per year, saving 58.8~TiB of storage. At Amazon S3 Infrequent Access pricing, this would cost \$9,031 per year, justifying the savings. Additionally, the storage savings will recur, year over year, while the capital expense of the Xeon for a year will be much less than \$9,000 and will depreciate only once.

\subsection{Safety Mechanisms}
\label{sec:safety}
During the initial roll-out, after \Lepton{} was activated for several days after April 14th, all of the several hundred TiB of compressed images had been downloaded in compressed form and decompressed twice in a row, once with a gcc, asan-enabled, \Lepton{} and another time with the default productionized icc \Lepton{} in multithreaded mode.

There are also alerts in place that page a member of the \Lepton{} team if a particular chunk is unable to be decompressed. The construction of this alert required some care(\S~\ref{sec:timebound}).  There is a ``playbook entry'' for
the person on call to immediately disable \Lepton{}.

The shutoff switch operates by placing a file with a predetermined name in /dev/shm, and the \Lepton{} system checks for that file before compressing new chunks. Most \Dropbox{} configuration files take between 15 and 45 minutes to fully deploy, but this mechanism allows a script to populate the file across all hosts that encode \Lepton{} within 30 seconds.

The blockservers also never admit chunks to the storage system that fail to round-trip---meaning, to decode identically to their input. Additionally, all memory pages holding compressed data in memory are protected at the OS level before the round-trip test, and an md5sum is done of the initial compressed file to be compared with the result stored on disk, so the memory contents cannot change due to a user-level bug.  Corruptions in compressing will be detected immediately, and the files will be compressed using Deflate instead.

For a file that fails to round-trip, it can be difficult to distinguish
between a badly formed/unsupported JPEG file versus a real program bug,
but as long as the compressor and decompressor are deterministic, a small
level of such failures is acceptable.

For added safety, there is an automated verification process that searches for images that succeeded in a round-trip once but then fail
a subsequent round-trip test, or fail when decompressed with the address-sanitizing gcc build of \Lepton{}. If
either of those occur, a member of the \Lepton{} team is paged and the failing
data is saved. This process has currently gone through over
four billion files and has caused four alerts (\S~\ref{sec:false_alarms}).

During roll-out of a new \Lepton{} version it will be ``qualified''
using the automated process over a billion images.
Additionally it must be able to decompress another billion images already compressed in the store.
Currently a candidate which fails to do so also causes the \Lepton{} team to be paged. This alert has never triggered.

For the first two weeks of ramp-up, the system was completely reversible. Every
chunk uploaded with \Lepton{} was concurrently also uploaded to a separate S3
bucket (the ``safety net'') with the standard Deflate codepath.
This means that in the worst case, requests could fail-over directly to the
safety net until the affected files were repaired.

Before enabling \Lepton{}, the team did a mock disaster recovery training
(DRT) session where a file in a test account was intentionally corrupted and
recovered from the safety net.
However, we never needed to use this mechanism to recover any real user files.

We have since deleted the safety net and depend on other controls to keep \Lepton{} safe. Our rationale for this was that uploading to a separate bucket causes a performance degradation since all images would upload in the max of latency between \Dropbox{} datacenters and S3, plus associated transaction and storage fees. We may re-enable the safety net during future format upgrades.

Even with the safety net disabled, we believe there are adequate recovery plans in place in case of an
unexpected error.
Every file that has been admitted to the system with \Lepton{} compression has also round-tripped at least once in order to be admitted.
That means that a permanent corruption would expose a hypothetical nondeterminism in the system. But it also means that if the same load/perf circumstances were recreated, the chunk would probably be decodable again with some probability, as it was decoded exactly correctly during the original round-trip check. Thus, with sufficient retries, we would expect to be able to recover the data.  That said, it would be a significant problem if there were a nondeterminism in the \Lepton{} system. After 4 billion successful determinism tests, however, we believe the risk is as small as possible.

\section{Anomalies at Scale}
\label{s:casestudies}

With a year of \Lepton{} operational experience, there have been a number of anomalies encountered and lessons learned. We share these in the hopes that they
will be helpful to the academic community in giving context about challenges
encountered in the practical deployment of format-specific compression tools in an exabyte-scale network filesystem.

\subsection{Reversed indices, bounds checks and compilers}
\label{sec:reversedindices}
During the very first qualification of 1 billion files, a handful of images passed
the multithreaded icc-compiled check, but those images would
  occasionally fail the gcc roundtrip check with a segfault. The stack trace revealed the
multidimensional statistic-bin index computation was reversed.
If deployed, this would have required major backwards compatibility
contortions to mimic the undefined C++ behavior as compiled with icc;
bins would need to be aliased for certain versions of \Lepton{}.

In response to this discovery, the statistic bin was abstracted with a class
that enforced bounds checks on accesses. Consequently, the duration
of encodes and decodes are 10\% higher than they could be, 
but bounds checks help guard against undefined behavior.

\subsection{Error codes at scale}

This table shows a variety of exit codes that we have observed during the first 2 months of backfill.

%
%
%
%
%
%
%
%
%
%
%
%
%
%
%
%
%
%
%
%
%
%
%

  \footnotesize
\noindent
\begin{tabular}{@{}c@{\hspace{12pt}}c@{}}
\begin{tabular}{@{}p{80pt}<{\dotfill}@{}r@{}}
Success & 94.069\% \\
Progressive & 3.043\% \\
Unsupported JPEG & 1.535\% \\
Not an image & 0.801\% \\
4 color CMYK & 0.478\% \\
$>$24 MiB mem decode & 0.024\% \\
$>$178 MiB mem encode & 0.019\% \\
Server shutdown & 0.010\% \\
\end{tabular}
&
\begin{tabular}{@{}p{80pt}<{\dotfill}@{}l@{}}
``Impossible'' & 0.006\% \\
Abort signal & 0.006\% \\
Timeout & 0.004\% \\
Chroma subsample big & 0.003\% \\ %
AC values out of range & 0.001\% \\
Roundtrip failed & 0.001\% \\
OOM kill & $10^{-5}$\% \\
Operator interrupt & $10^{-6}$\% \\ %
\end{tabular}
\end{tabular}
\paperfontsize{}

The top 99.9864\% of situations were anticipated: from graceful shutdown, to deciding not to encode JPEG files that consist entirely of a header, to unsupported Progressive and CMYK JPEGs, and chroma subsampling that was larger than the slice of framebuffer in memory.

The \Lepton{} program binary could process these types of images, e.g., by
allocating more memory, an extra model for the 4th color channel, or
sufficient memory on decode to keep the progressive image resident. However, for simplicity, these features were
intentionally disabled in \Dropbox{} as they account for a small
fraction of files.

Some codes were unexpected, e.g., incorrect thread protocol communication
(listed as ``Impossible''), or ``Abort signal'', since {\tt SECCOMP} disallows
\texttt{SIGABRT}. By contrast, a small level of ``Roundtrip failed'' was expected,
largely because of corruptions in the JPEG file (e.g., runs of zeroes written by
hardware failing to sync a file) that cannot always be represented in the \Lepton{}
file format.

\subsection{Poor p95 latency from Huge Pages}
During the \Lepton{} roll-out, after the qualification round, a
significant fraction of machines had significantly higher average and
p99 latency, and could take 2--3$\times$ as long as the isolated
benchmarks. It was even possible to have a decode take
30 seconds to even begin processing. The time would elapse before a
single byte of input data was read. Reboots could sometimes
alleviate the issue, but on the affected machines it would come back. Also,
when services on the machine were halted and benchmarks run in
isolation, the problem disappeared altogether and the machine performed
as expected.

On affected machines, performance counters attributed
15--20\% of the time to the kernel's page-table routines.

\footnotesize
\begin{tabular}{@{\hspace{3pt}}llll@{}}
9.79\%  & $[kernel]$                  &$[k]$&isolate\_migratepages\_range \\
4.04\%  & $[kernel]$                 &$[k]$&copy\_page\_range \\
3.16\%  & $[kernel]$                 &$[k]$&set\_pfnblock\_flags\_mask \\
\end{tabular}
\paperfontsize{}

\begin{figure}
\includegraphics[width=7.95cm]{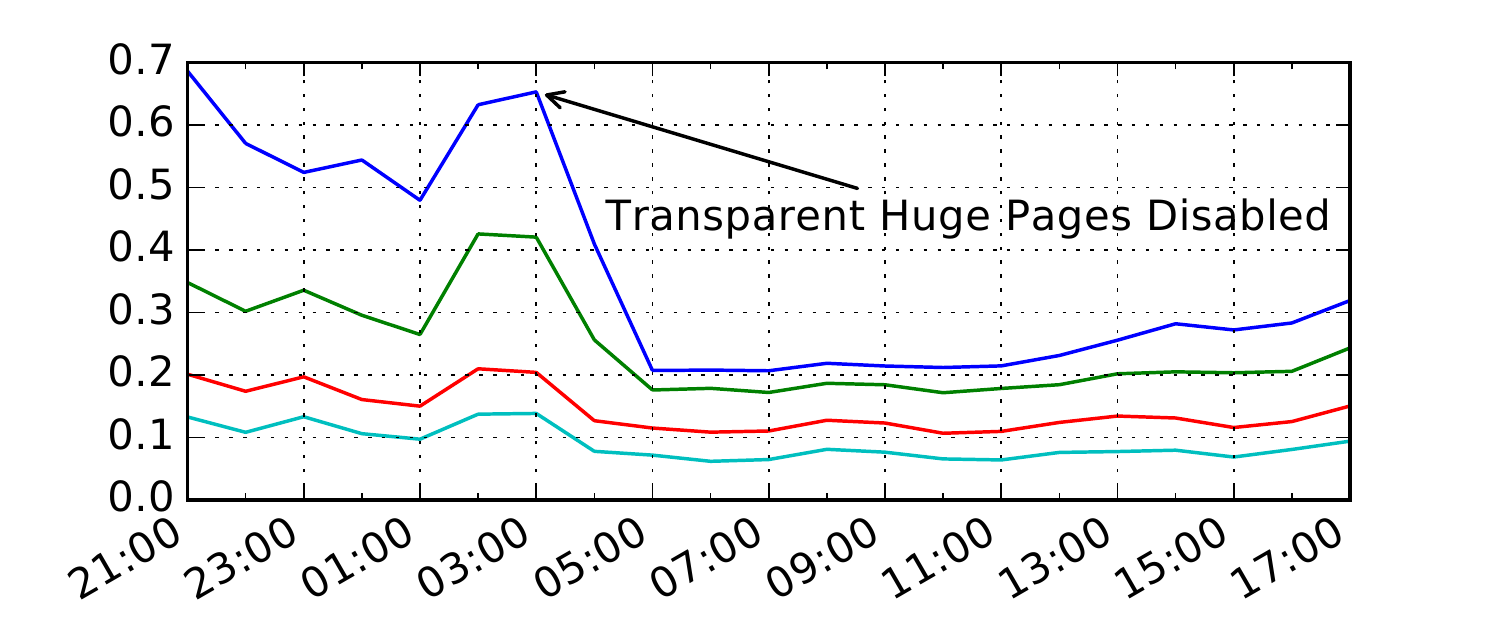}
\caption{Hourly p99/p95/p75/p50 latency for decodes. Transparent huge pages disabled April 13 at 03:00.}
\label{fig:thp}
\end{figure}

These kernel routines implicated transparent huge pages (THP), and
each affected machine had THP enabled $[Always]$, but
most unaffected machines had them disabled. Disabling THP solved
the issue (Figure~\ref{fig:thp}).

Additionally, when THP is enabled, Linux continuously
defragments pages in an attempt to build a full 2~MiB page of free
memory for an application requesting large ranges of data.  Since
\Lepton{} requests 200 MiB of space at initialization time, with no
intent to use more than 24 MiB for decodes, Linux may prepare a
significant number of huge pages for use, causing the process to be
blocked during defragmentation. These pages are consumed without penalty over the next 10 decodes, meaning that the p95 and p99 times are disproportionately affected by the stall (compared with the median times).

\subsection{Boiling the frog}

\begin{figure}
\includegraphics[width=7.95cm]{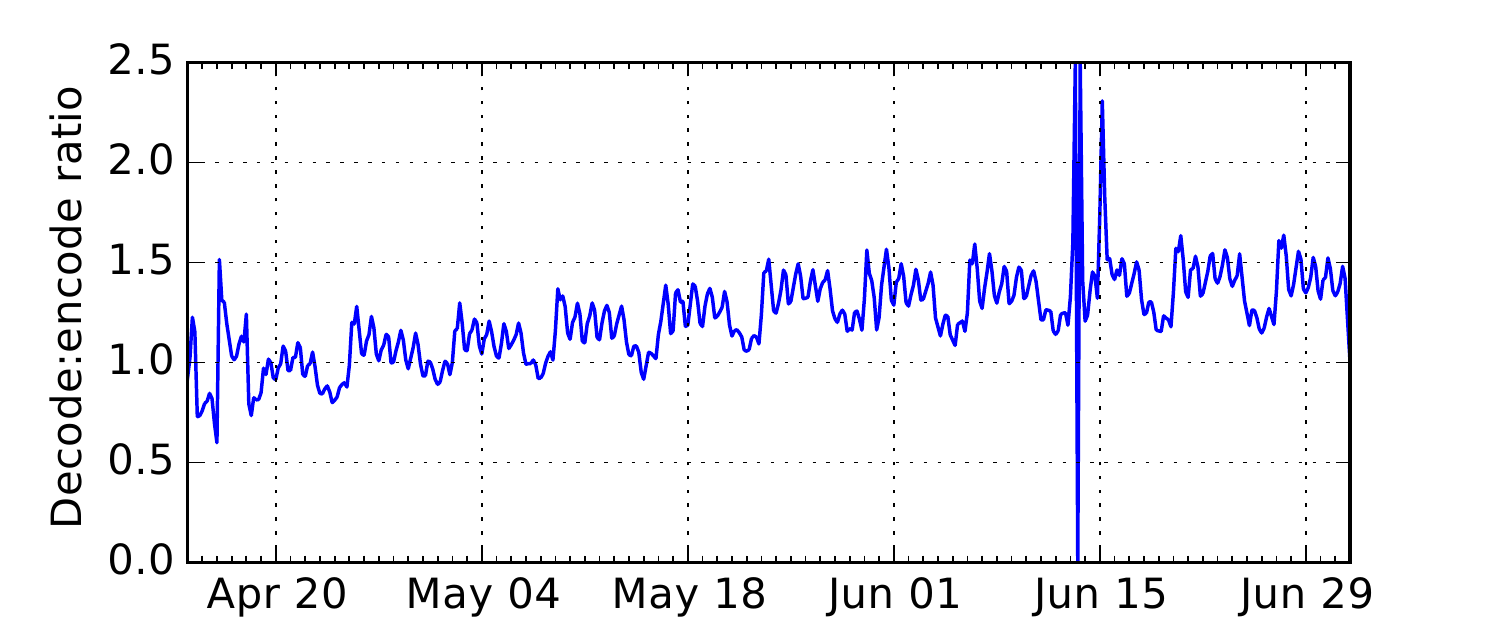}
\caption{\Lepton{} decode:encode ratio on the serving path.}
\label{fig:encode_ratio_6mo}
\end{figure}

Currently, the \Lepton{} system decodes about 1.5$\times$ to twice as many files as it encodes.
However,
during the initial roll-out, months before the backfill system, the ratio of decodes to encodes was much less than 1.0, since each old
photo was compressed using Deflate, not
\Lepton{}, and only new photos need a \Lepton{} decompress.  This can be
seen in the historical graph of the decode:encode ratio over time in
Figure \ref{fig:encode_ratio_6mo}. Akin to ``boiling the frog,'' it was
not obvious that the actual hardware requirements would be significantly
higher than those needed from having reached 100\% of users in the first
weeks.

\begin{figure}
\includegraphics[width=7.95cm]{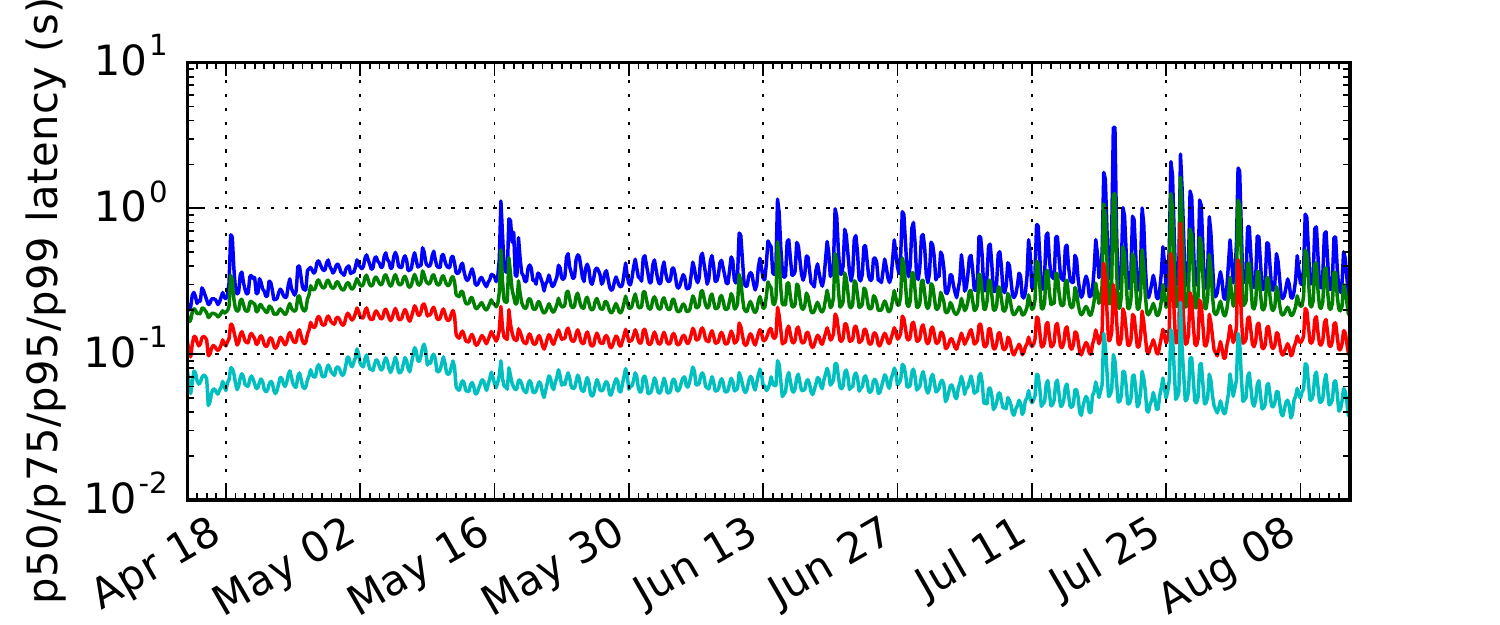}
\caption{Decode timing percentiles, starting with the roll-out and
  building up over months to multi-second p99s.}
\label{fig:boiling_the_frog}
\end{figure}

To react to these new requirements for decodes, we built the outsourcing
system (\S~\ref{sec:outsourcing}). But until that
system rolled out, for several months, at peak, our 99th-percentile decode time was
in the seconds, as seen in Figure \ref{fig:boiling_the_frog}.

\subsection{\Dropbox{} \CameraUpload{} degradation}
Before removal of the safety net, each image would be uploaded
compressed to the \Dropbox{} store and uncompressed to the S3 safety
net. During maintenance in one East-Coast datacenter, each top of rack switch
required a reboot. Traffic was rerouted to another datacenter. The transition was going
well, but on June 13 at 8:40:25, once most traffic
had moved to the new location, S3 ``put'' operations began to fail
sporadically from truncated uploads. The safety-%
net feature was writing more data to S3 from the new location than \emph{all of the rest of \Dropbox{} combined}, and the capacity of our S3 proxy machines was
overtaxed by the safety-net mechanism.

For uploads, the availability
dropped to 94\% for the 9 minutes required to diagnose the situation,
and \CameraUpload s from phones were disproportionately affected, as mobile devices are a very common means for users to capture images. Availability of this service dropped to 82\%,
since each photograph upload required a write to the safety net.
Once the situation was identified, \Lepton{} encodes were disabled in 29
seconds, using the shutoff
switch (\S~\ref{sec:safety}) in /dev/shm, stopping
extra traffic to S3. Traffic returned to normal at 8:49:54.

An irony emerged: a system we designed as a belt-and-suspenders safety net ended up causing our users trouble, but has never helped
to resolve an actual problem.

\subsection{Decodes that exceed the timeout window}
\label{sec:timebound}
 With thousands of servers decoding chunks, there are often unhealthy
 systems that are swapping, overheating, or broken.  These can become
 stuck during a \Lepton{} decode and time out.
 Because these events happen regularly, they must be investigated automatically without involving a human operator.%

Instead, any decode exceeding a timeout is uploaded to an S3 queue bucket. Chunks in
this queue bucket are decompressed on an isolated, healthy cluster
without a timeout using the gcc-asan as well as the icc build of \Lepton{}. If the chunk is successfully decoded 3
times in a row with each build, then the chunk is deleted from the
bucket. If any of those decodes fails, a human is signaled.

\subsection{Alarm pages}
\label{sec:false_alarms}
As of this submission, anomalies in the \Lepton{} system have
caused an on-call engineer to be paged four times.

\paragraph{Assert failed in sanitizing build only.} The first alarm occurred just days after \Lepton{}
was activated for 0.1\% of users, on April 8. When reconstructing
the Huffman coded data, each thread asserts that the number of bytes
produced matches the number of bytes decoded on the initial
compression. A file tripped this assert in the gcc-asan build that was
disabled for the icc production build, so the icc build admitted the file.

The solution was to compile \Lepton{} with all meaningful asserts
enabled and to check whether any of the existing 150~TiB of images
tripped the assert. Luckily no other files triggered the assert.
We deployed stricter code that will not admit such files.

\paragraph{Single- and multi-threaded code handled corrupt JPEGs differently.} The second alarm was triggered on May 11 because of a bug in the single-threaded
code.  The single-threaded decoder wrote all output directly to the
file descriptor, whereas in the multithreaded decoder, each thread wrote
to a fixed sized memory area.  When the JPEG was
sufficiently corrupt, the size would be incorrectly computed, but the
writes to the memory area would be truncated in multi-threaded mode, yet
the direct writes to the stream would be unbounded in single-thread
mode. The fix was to make sure single-threaded decodes bounded their
writes to the stream as if it were a fixed memory region.

\paragraph{After open-source release, fuzzing found bugs
in parser handling of corrupt input.} The third alarm was caused by a security researcher~\LeptonCVE{}, who fuzzed the open-source release of \Lepton{} and found bugs
in the
\mbox{uncmpjpg} JPEG-parsing library that \Lepton{} uses. The library did not
validate that the Huffman table had sufficient space for the
data. Uncmpjpg would overwrite global memory past the array with
data from the untrusted input.  A similar bug existed in uncmpjpg's
quantization table index, which also opened up a buffer
overrun. The response was to replace every raw array with a
bounds-checked {\tt std::array}, to avoid similar attacks in the
future. It was unfortunate that we did not apply this philosophy after
the earlier ``reversed indices'' incident (\S~\ref{sec:reversedindices}). Fortunately,
the deployed system is protected with {\tt SECCOMP}, preventing escalation of
privileges.

\paragraph{Accidental deployment of incompatible old version.}
The final alarm was the result of a series of operational mistakes. On Dec.~12, 2016, a new
team member was deploying Lepton on some blockservers. The
internal deployment tool asks the operator to specify the hash
of a \Lepton{} build to deploy. These builds have all been
``qualified,'' meaning they successfully compressed and then
decompressed a billion JPEGs with both optimized and
sanitizing decoders, yielding identical results to the input.

Our historical practice has been to retain earlier ``qualified''
builds as eligible for deployment, so that \Lepton{} can be rolled
back if necessary. However, because \Lepton{}'s file format has
evolved over time, the earliest qualified builds are not
compatible with more recent versions. When features were added,
an older decoder may not be able to decode a newer
file. When \Lepton{}'s format was made stricter,
an older encoder may produce files that are rejected by a newer
decoder. At the time
of such upgrades, we searched for and re-encoded JPEG files in
\Dropbox{} as necessary, but we did not remove the older software from
the list of qualified builds.

Typically, our team members deployed \Lepton{} to blockservers by
specifying the hash of the most recent qualified build in the
deployment tool. However, our
documentation did not properly inform the new employee of this
practice, and they simply left the field blank. This caused the
deployment tool to use an internal default value of the hash,
which had been set when \Lepton{} was first deployed and never
updated. As a result, the very first qualified version of \Lepton{}
was accidentally deployed on some blockservers.

The first warning sign was availability dropping to 99.7\% for upload and download endpoints. This was due to the oldest
qualified \Lepton{} code being unable to decode some newly compressed images because of
minor additions to the format. An additional alarm was triggered
after other blockservers (ones that did not receive the bad
configuration change) found themselves unable to decode some files
that had been written by blockservers that did receive the change.

As operators were occupied trying to roll back the configuration
change, it took two hours before \Lepton{} was disabled, during which
time billions of files were uploaded. We performed a scan over all
these files, decoding and then re-encoding them if necessary into the
current version of the \Lepton{} file format. Ultimately, 18 files had to be re-encoded.%

This was an example of a number of procedures gone wrong. It confirms
the adage: we have met the enemy and he is us. The incident has caused us to reconsider whether a
``qualified'' version of \Lepton{} ought to remain eternally qualified
for deployment, the behavior and user interface of the deployment
tools, and our
documentation and onboarding procedures for new team members.

\section{Limitations and Future Work}
\label{sec:future_work}

\Lepton{} is currently deployed on \Dropbox{}'s back-end file servers, and is
transparent to client software. In the future, we intend to move the
compression and decompression to client software, which will save 23\%
in network bandwidth when uploading or downloading JPEG images.

\Lepton{} is limited to JPEG-format files, which account for roughly
35\% of the \Dropbox{} filesystem. Roughly another 40\%
is occupied by H.264 video files, many of which are encoded by
fixed-function or power-limited mobile hardware that does not use
the most space-efficient lossless compression methods.
We intend to explore
the use of \Lepton{}-like recompression for mobile video files.

\section{Conclusion}
\Lepton{} is an open-source system that compresses JPEG images by 23\%
on average. It has been deployed on the production \Dropbox{} network
filesystem for a year and has so far compressed more than 150
billion user JPEG files that accounted for more than 203 PiB.
\Lepton{} was designed to be deployable on a distributed
file-serving backend where substrings of a file must be decodable
independently, with low time-to-first-byte and time-to-last-byte. The
system demonstrates new tradeoffs between speed, compression
efficiency, and deployability in the context of a large-scale
distributed filesystem back-end.

In a year of production use and hundreds of billions of downloads,
deployment has been relatively smooth. We have never been unable
to decode a stored file. The issues we have encountered
have involved human error and procedural failures, non-obvious ways in which the system created load
hotspots, and difficulties in ensuring
deterministic behavior from a highly optimized C++ program processing
untrusted input from diverse sources. We have shared a number of
deployment case studies and anomalies in the hopes that they will be
helpful to the academic community in giving context about challenges
encountered in the practical deployment of format-specific compression
tools at a large scale.

\section{Acknowledgments}

We would like to thank the reviewers and NSDI program committee for their detailed
comments and suggestions.

This work was funded by Dropbox. DH, KE, and CL are Dropbox employees.
KW's participation was as a paid consultant and was not part of his Stanford
duties or responsibilities.

We would like to thank Jongmin Baek, Sujay Jayakar, and Mario Brito for
their ideas on the Lepton approach, and Preslav Le,
David Mah, James Cowling, Nipunn Koorapati, Lars Zornes, Rajat Goel,
Bashar Al-Rawi, Tim Douglas, and Oleg Guba for their assistance and guidance on
deploying and rolling out Lepton to Dropbox infrastructure.

This project would not have been possible without the continuous
support and encouragement from Dropbox leadership including
Dimitry Kotlyarov, Nahi Ojeil, Bean Anderson, Ziga Mahkovec, David Mann,
Jeff Arnold, Jessica McKellar, Akhil Gupta, Aditya Agarwal, Arash
Ferdowsi, and Drew Houston.

\label{LastPage}
{
\newlength{\bibitemsep}\setlength{\bibitemsep}{.2\baselineskip plus .05\baselineskip minus .05\baselineskip}
\newlength{\bibparskip}\setlength{\bibparskip}{0pt}
\let\oldthebibliography\thebibliography
\renewcommand\thebibliography[1]{%
  \oldthebibliography{#1}%
  \setlength{\parskip}{\bibitemsep}%
  \setlength{\itemsep}{\bibparskip}%
}
\setlength{\bibitemsep}{.2\baselineskip plus .05\baselineskip minus .05\baselineskip}

\bibliographystyle{acm}
\bibliography{ms}

\begin{thebibliography}{10}

\bibitem{leptonrepo}
Lepton source repository.
\newblock \leptonrepo{}.

\bibitem{mozjpeg}
{\sc Aas, J.}
\newblock Introducing the `mozjpeg' project, 2014.
\newblock
  \url{https://blog.mozilla.org/research/2014/03/05/introducing-the-mozjpeg-project/}.

\bibitem{brotli}
{\sc Alakuijala, J., and Szabadka, Z.}
\newblock Brotli compressed data format.
\newblock {RFC} 7932, {RFC Editor}, July 2016.
\newblock \url{https://tools.ietf.org/html/rfc7932}.

\bibitem{rfc6386}
{\sc Bankoski, J., Koleszar, J., Quillio, L., Salonen, J., Wilkins, P., and Xu,
  Y.}
\newblock {{VP8} Data Format and Decoding Guide section 13.2}.
\newblock {RFC} 6386, {RFC Editor}, November 2011.
\newblock \url{https://tools.ietf.org/html/rfc6386#section-13.2}.

\bibitem{zstd}
{\sc Collet, Y., and Turner, C.}
\newblock Smaller and faster data compression with {Zstandard}, 2016.
\newblock
  \url{https://code.facebook.com/posts/1658392934479273/smaller-and-faster-data-compression-with-zstandard/}.

\bibitem{deflate}
{\sc Deutsch, P.}
\newblock {DEFLATE} compressed data format specification version 1.3.
\newblock {RFC} 1951, {RFC Editor}, May 1996.
\newblock \url{https://tools.ietf.org/html/rfc1951}.

\bibitem{cve2016_6234}
{\sc Grassi, M.}
\newblock Some memory corruptions in lepton.
\newblock \url{https://github.com/dropbox/lepton/issues/26}.

\bibitem{huffman}
{\sc Huffman, D.~A.}
\newblock A method for the construction of minimum-redundancy codes.
\newblock {\em Proceedings of the IRE 40}, 9 (1952), 1098--1101.

\bibitem{ijg}
{Independent JPEG Group}.
\newblock \url{http://www.ijg.org/}.

\bibitem{jpeg}
{\em Information technology --- Digital compression and coding of
  continuous-tone still images: Requirements and guidelines}, September 1992.
\newblock ITU-T Rec. T.81 and ISO/IEC 10918-1:1994
  (\url{https://www.w3.org/Graphics/JPEG/itu-t81.pdf}).

\bibitem{lakhani}
{\sc Lakhani, G.}
\newblock {DCT} coefficient prediction for {JPEG} image coding.
\newblock In {\em 2007 IEEE International Conference on Image Processing\/}
  (San Antonio, TX, USA, Sept 2007), vol.~4, pp.~IV--189 to IV--192.

\bibitem{paq}
{\sc Mahoney, M.}
\newblock Paq: Data compression programs, 2009.
\newblock \url{http://mattmahoney.net/dc/}.

\bibitem{jpgrescan}
{\sc Merritt, L.}
\newblock {JPEGrescan}: losslessly shrink any {JPEG} file, 2013.
\newblock \url{https://github.com/kud/jpegrescan}.

\bibitem{tworandomchoices}
{\sc Mitzenmacher, M., Richa, A.~W., and Sitaraman, R.}
\newblock The power of two random choices: A survey of techniques and results.
\newblock In {\em Handbook of Randomized Computing\/} (2000), Kluwer,
  pp.~255--312.

\bibitem{lzma}
{\sc Pavlov, I.}
\newblock {LZMA SDK}, 2007.
\newblock \url{http://www.7-zip.org/sdk.html}.

\bibitem{asan}
{\sc Serebryany, K., Bruening, D., Potapenko, A., and Vyukov, D.}
\newblock {AddressSanitizer}: A fast address sanity checker.
\newblock In {\em Proceedings of the 2012 USENIX Conference on Annual Technical
  Conference\/} (Boston, MA, USA, 2012), USENIX ATC'12, USENIX Association,
  pp.~28--28.

\bibitem{packjpg}
{\sc Stirner, M., and Seelmann, G.}
\newblock Improved redundancy reduction for {JPEG} files.
\newblock In {\em Proceedings of the 2007 Picture Coding Symposium\/} (Lisbon,
  Portugal, 2007).

\bibitem{teuhola1978compression}
{\sc Teuhola, J.}
\newblock A compression method for clustered bit-vectors.
\newblock {\em Information processing letters 7}, 6 (1978), 308--311.

\bibitem{tinyjpg}
{\sc {Voormedia B.V.}}
\newblock {TinyJPG}.
\newblock \url{https://tinyjpg.com/}.

\bibitem{zigzag}
{\sc Wallace, G.~K.}
\newblock The {JPEG} still picture compression standard.
\newblock {\em IEEE Transactions on Consumer Electronics 38}, 1 (1992), 34.

\bibitem{undefined-behavior}
{\sc Wang, X., Chen, H., Cheung, A., Jia, Z., Zeldovich, N., and Kaashoek,
  M.~F.}
\newblock Undefined behavior: what happened to my code?
\newblock In {\em Proceedings of the Asia-Pacific Workshop on Systems\/}
  (Seoul, Republic of Korea, 2012), ACM, p.~9.

\bibitem{xu2016context}
{\sc Xu, X., Aichta, Z., Govindan, R., Lloyd, W., and Ortega, A.}
\newblock Context adaptive thresholding and entropy coding for very low
  complexity {JPEG} transcoding.
\newblock In {\em 2016 IEEE International Conference on Acoustics, Speech and
  Signal Processing (ICASSP)\/} (Shanghai, China, 2016), IEEE, pp.~1392--1396.

\bibitem{lz77}
{\sc Ziv, J., and Lempel, A.}
\newblock A universal algorithm for sequential data compression.
\newblock {\em IEEE Transactions on Information Theory 23}, 3 (1977), 337--343.

\end{thebibliography}
}
\appendix
\section{Appendix}
\subsection{File Format}

\begin{tabular}{|c|}
\hline
Magic Number (0xcf, 0x84) \hspace{.7cm}(2 bytes)\\
\hline
Version (0x01)  \hspace{2.65cm}(1 byte)\\
\hline
Skip serializing header? ( Y $\|$ \rz~) (1 byte) \\
\hline
Number of Thread Segments \hspace{.4cm} (4 bytes) \\
\hline
Truncated Git Revision \hspace{1.1cm} (12 bytes) \\
\hline
Output File Size \hspace{2.25cm}(4 bytes) \\
\hline
Zlib Data Size \hspace{2.5cm} (4 bytes) \\
\hline
\cellcolor{gray!25}Zlib Data \\
\cellcolor{gray!25}\\
\cellcolor{gray!25}\begin{tabular}{|c|}
\hline
\cellcolor{white!25}JPEG Header Size (4 bytes) \\
\hline
\cellcolor{white!25}JPEG Header \\
\hline
\cellcolor{white!25}Pad Bit ( 0 $\|$ 0xFF ) \\
\hline
\cellcolor{white!25}Per-Thread Segment Information \\
\cellcolor{white!25}\begin{tabular}{|cc|}
\hline
\cellcolor{gray!12}Thread Segment Vertical Range &\cellcolor{gray!12}(2 bytes) \\
\hline
\cellcolor{gray!12}Size of Thread Segment Output &\cellcolor{gray!12}(4 bytes) \\
\hline
\cellcolor{gray!12}Huffman Handover Word &\cellcolor{gray!12}(2 bytes) \\
\hline
\cellcolor{gray!12}DC per channel &\cellcolor{gray!12}(8 bytes) \\
\hline
\end{tabular} \\
\cellcolor{white!25}\\
\hline
\cellcolor{white!25}Number of RST markers \\
\hline
\cellcolor{white!25}Total Number of 8x8 JPEG Blocks per channel\\
\hline
\cellcolor{white!25}Arbitrary data to prepend to the output \\
\hline
\cellcolor{white!25}Arbitrary data to append to the output \\
\hline
\end{tabular}
\cellcolor{gray!25}\\
\cellcolor{gray!25}\\
\hline
Interleaved Arithmetic Coding Section \\
\\
\begin{tabular}{|c|}
\hline
\cellcolor{gray!12}Thread Segment Id \\
\hline
\cellcolor{gray!12}Length ($256 \| 4096 \| 65536 \|$arbitrary) \\
\hline
\cellcolor{gray!12}Arithmetic coded data \\
\hline
\hline
\cellcolor{gray!12}Thread Segment Id \\
\hline
\cellcolor{gray!12}Length ($256 \| 4096 \| 65536 \|$arbitrary) \\
\hline
\cellcolor{gray!12}Arithmetic coded data \\
\hline
\end{tabular} \\
... \\
May be repeated many times per thread segment \\
... \\
\end{tabular}

\subsection{Using prior information to predict and encode DCT coefficients}
\label{sec:bin_index}
Each 8x8 JPEG block has 49 2D DCT coefficients, 14 1D DCT coefficients, and one
DC coefficient (\S~\ref{sec:dct_coefficient_priors}).
Lepton encodes each kind of coefficient using the same Exp-Golomb code (unary
exponent, then sign bit, then residual bits) but with different methods for
indexing the adaptive arithmetic code's bins.
Prior information, such as neighboring blocks, is used to predict bin indices;
higher correlation between the predicted indices and the actual coefficient
values yields better compression ratios.

\subsubsection{Predicting the 7x7 AC coefficients}
Lepton first encodes the number of non-zero coefficients in the block,
$n \in \{0,\dots,49\}$, by emitting 6 bits. Since the number of non-zero
coefficients in the above and left blocks approximately predicts $n$, the bins
are indexed by
$\left\lfloor \log_{1.59} \left( \frac{n_A + n_L}{2} \right) \right\rfloor \in \{0,\dots,9\}$.
The bin for each bit is further indexed by the previously decoded bits, so that
the total number of bins (for encoding $n$) is $10 \times (2^6-1)$.

\begin{figure}
\includegraphics[width=7.95cm]{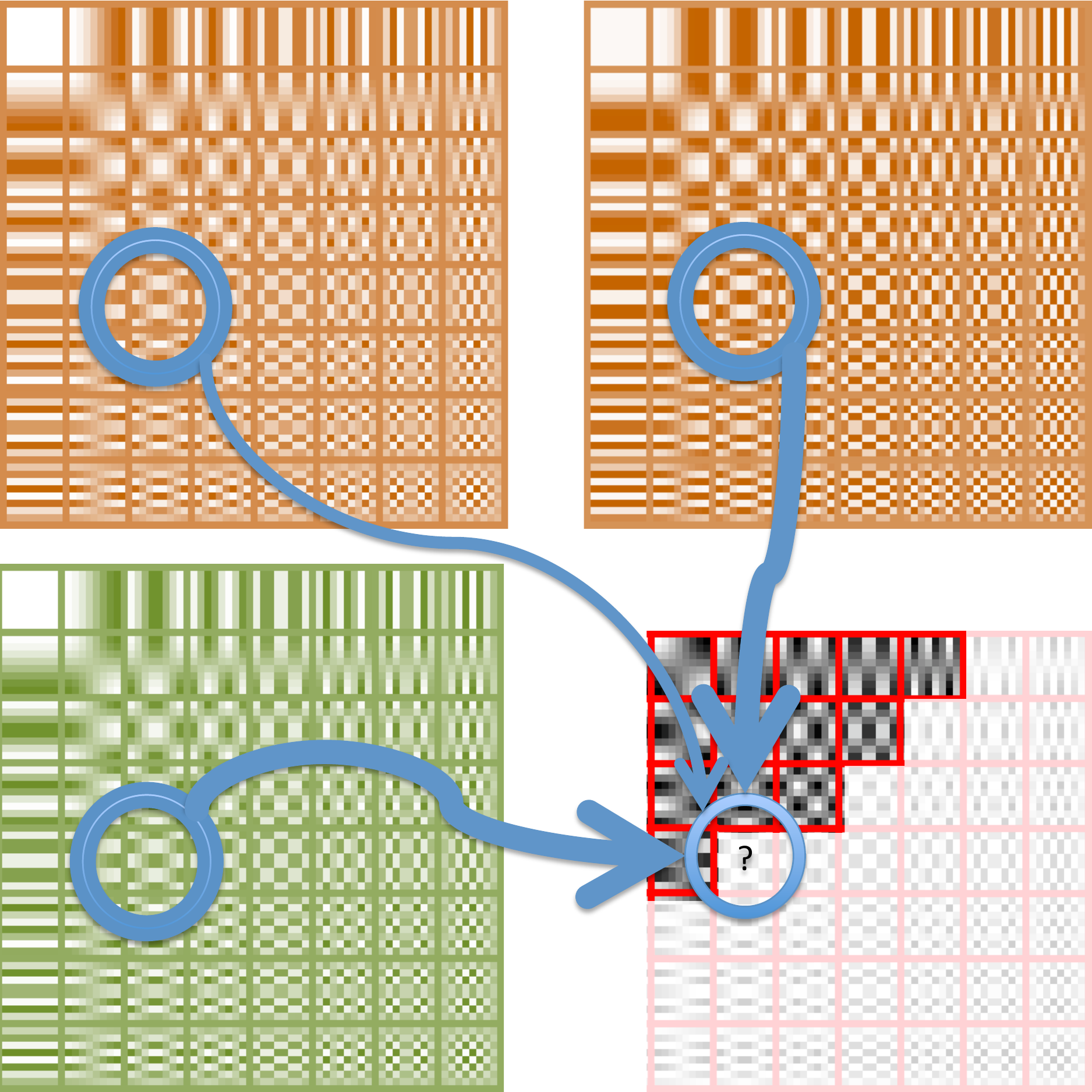}
\caption{The neighboring coefficients at the same coordinates are averaged to
predict the 7x7 coefficient.}
\label{fig:7x7_priors}
\end{figure}

The 7x7 coefficients are encoded in zigzag order~\cite{zigzag}, which yields
a 0.2\% compression improvement over raster-scan order.
For each coefficient~$F$, we compute a weighted average of the corresponding
coefficient from the above, left, and above-left blocks
(Figure~\ref{fig:7x7_priors}):
$\bar{F} = \frac{1}{32}\left( 13 F_A + 13 F_L + 6 F_{AL} \right)$.
Each coefficient is Exp-Golomb encoded using bins indexed by
$\left( \left\lfloor \bar{F} \right\rfloor,
        \left\lfloor \log_{1.59} n \right\rfloor \right)$.
Each time a non-zero coefficient is encoded, $n$ is decremented; the block is
finished when $n=0$.

\subsubsection{Predicting the 7x1 and 1x7 AC coefficients}
\label{s:lakhani}

The 7x1 and 1x7 coefficients represent image variation purely in the horizontal
and vertical directions. Lepton encodes them similarly to the 7x7 coefficients,
but instead of predicting each coefficient using a weighted average, we use
a more sophisticated formula inspired by Lakhani~\cite{lakhani}. When encoding
the 1x7 coefficients, Lepton has already encoded the 7x7 coefficients, as well
as the full block to the left of the current block. We combine this prior
information with the additional assumption that the image pixel values are
continuous across the block edge --- i.e., that \( P_L(7,y) \approx P(0,y) \).

The block's pixels are defined to be a linear combination of orthonormal DCT
basis functions $B$:
\[
P(x,y) = \sum\limits_{u=0}^7 \sum\limits_{v=0}^7 B(x,u) B(y,v) F_{uv}
\]
Written as matrices, \( P = B^T F B \) where \( B^T B = 1 \).
The continuity assumption can then be rewritten as:
\begin{eqnarray*}
  e_7 B^T L B &\approx& e_0 B^T F B \\
  e_7 B^T L   &\approx& e_0 B^T F
\end{eqnarray*}
\begin{figure}
\includegraphics[width=7.95cm]{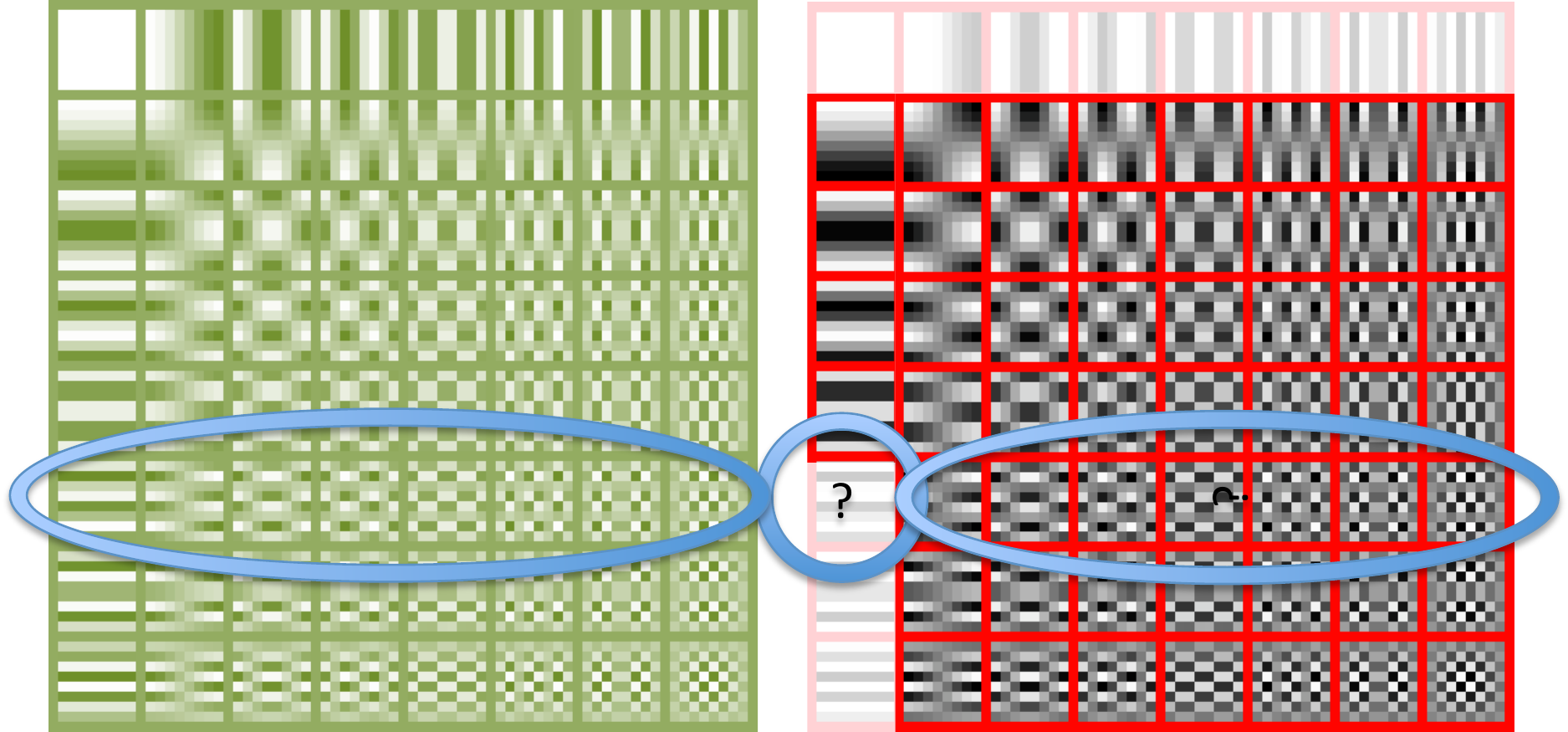}
\caption{An entire row (or column) of coefficients may be used to predict an edge DCT coefficient.}
\label{fig:7x1_priors}
\end{figure}
The left side is fully known from the block to the left, while the right side
is a linear combination of the known 7x7 coefficients and the unknown 1x7
coefficients, as shown in Figure~\ref{fig:7x1_priors}.
\[
  \sum\limits_{u=0}^7 B_{7u} L_{uv} \approx B_{00} F_{0v} + \sum\limits_{u=1}^7 B_{0u} F_{uv}
\]
We solve for the unknowns to predict $F_{0v}$:
\[
  \bar{F}_{0v} = \frac{1}{B_{00}} \left( \sum\limits_{u=0}^7 B_{7u} L_{uv} - \sum\limits_{u=1}^7 B_{0u} F_{uv} \right)
\]
The predicted $\bar{F}_{0v}$ is quantized to 7 bits and concatenated with the
non-zero count as the bin index for encoding the true coefficient $F_{0v}$.

\subsubsection{Predicting the DC coefficient}
\label{s:dc}

With all 63 of the AC coefficients known, the last block element to be encoded
is the DC coefficient. Instead of encoding this value directly, Lepton instead
predicts a DC coefficient and encodes the delta (the DC error term) between the
true value and the prediction. To make this prediction, Lepton first computes
the 8x8 pixel block (up to the constant DC shift) from the known AC
coefficients using inverse DCT.

\begin{figure}
\includegraphics[width=7.95cm]{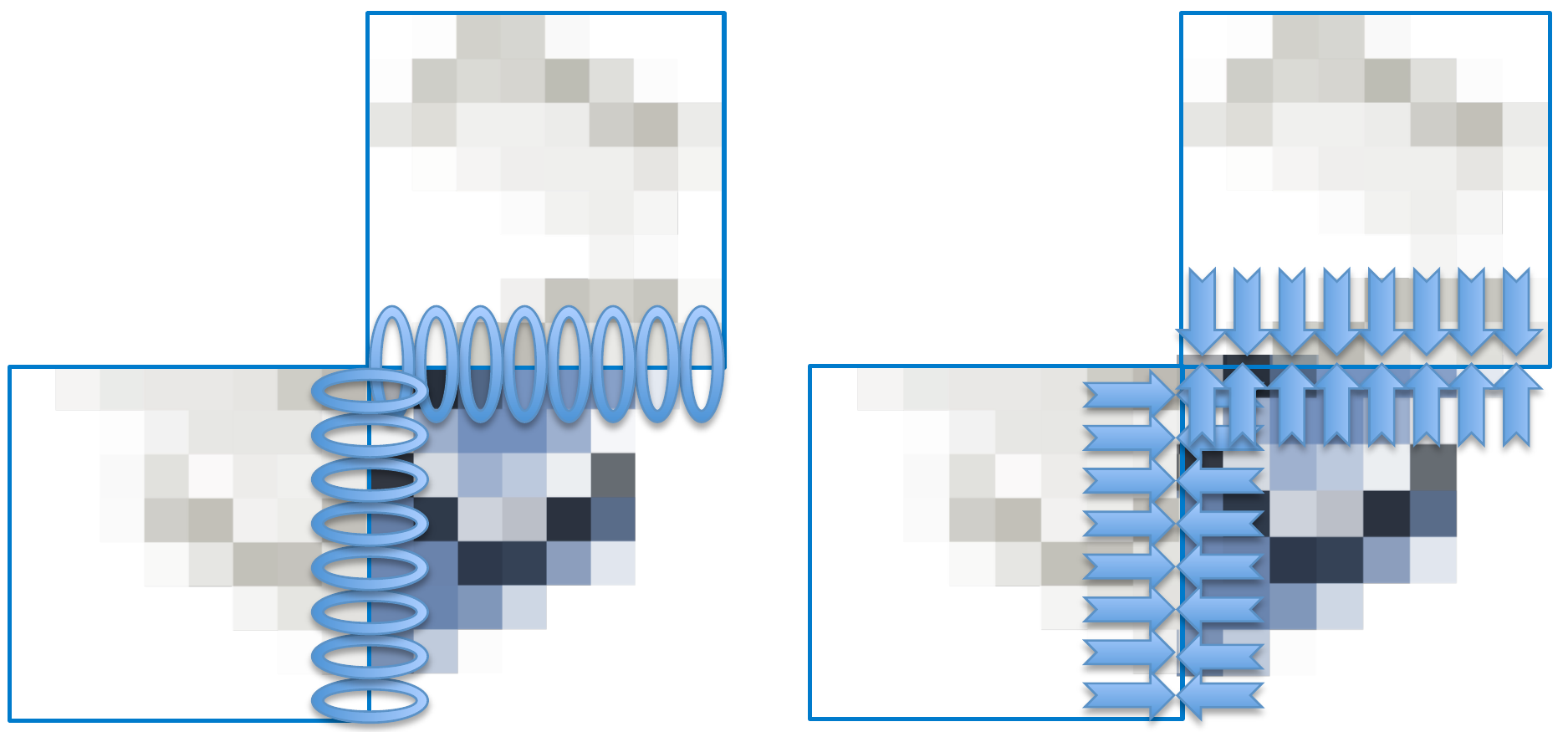}
\caption{Left: illustrates minimizing differences between pairs of pixels by
  predicting the DC value for the block being decoded (shaded in blue).
  DC adds a constant shift to all colors in the blue 8x8 block.
  Right: illustrates using gradients to interpolate colors between the blue
  block and its neighbors.}
\label{fig:dc_prior}
\end{figure}

A first-cut prediction, illustrated in Figure~\ref{fig:dc_prior} (left), might
be to compute the DC value that minimized the differences between all 16 pairs
of pixels at the borders between the current $8x8$ block and each of its left
and above neighbors. If we average the median 8 pairs and discard the 8 outlier
pairs, this technique compresses the DC values by roughly 30\% versus baseline
JPEG.

We can improve upon this prediction by observing that images tend to have
smooth gradients; for example, the sky fades from blue to orange towards the
horizon during a sunset.
Lepton interpolates the pixel gradients smoothly between the last two rows of
the previous block and the first two rows of the current block, as illustrated
in Figure~\ref{fig:dc_prior} (right). For each border pair, we predict the DC
value that would cause the two gradients to meet seamlessly. We finally encode
the DC error term between the true DC coefficient and the average of all 16
predictions. Lepton estimates the DC prediction confidence by subtracting the
maximum and minimum predictions (out of 16), and uses this value as the bin
index when Exp-Golomb encoding the error term. The combination of these
techniques yields another 10\% compression improvement over the first cut: the
final compression ratio is 40.1\% $\pm$ 8.7 better than the JPEG baseline.

\subsection{Common JPEG Corruptions}
\label{sec:corruptions}
In the \Lepton{} qualification process, there were several common JPEG anomalies that would trigger roundtrip failures.

Most prevalently, JPEG files sometimes contain or end with runs of zero bytes.  Likely these are caused by failures for an image editing tool or hard disk to sync pages to disk before a user depowered their machine. Many such images will successfully roundtrip with \Lepton{} since zero describes valid DCT data.

  However, RST markers foil this fortuitous behavior, since RST must be generated at regular block intervals in image space. Ironically the very markers that were designed to recover from partial corruption instead caused it for \Lepton{}.  For affected files, during these zero runs, the RST markers, beginning with a signature 0xff, would not be present. However, \Lepton{} blindly uses the RST frequencies in the header to insert them at regular intervals irrespective of the bytes in the original file. The fix for these zero-filled files regions at the end was to add a RST count to the \Lepton{} header, so that \Lepton{} could cease automatically inserting RST markers after the last one was recorded in the original file.

The RST solution did not fix issues with zero runs appearing in the middle of a scan, since one count cannot describe both the start and stop of RST insertions.  These corruptions manifest themselves as roundtrip failures.

A very common corruption was arbitrary data at the end of the file.  There are documented cases of cameras producing a TV-ready interlaced image file in a completely different format at the end of JPEG files produced. This let old cameras display images directly to TVs. One of the authors had a series of files where two JPEGs were concatenated, the first being a thumbnail of the second.  For these, the compression ratio is less, since \Lepton{} only applies to the thumbnail, but they do roundtrip.

Likewise, the JPEG specification does not mention whether partial bytes, filled with fewer than 8 bits, must be padded with zeros or with ones.  Most encoders pick a pad bit and use it throughout.  \Lepton{} stores this bit in the header.

\end{document}